\documentclass{emulateapj}
\slugcomment{To be published by the Astrophysical Journal}

\usepackage{natbib}

\usepackage{graphicx}
\usepackage{rotating}

\shorttitle{DM Halos and Bars: Collisionless Models}
\shortauthors{Villa-Vargas, Shlosman and Heller}

\begin{document}

\newcommand{\ps}{$\Omega_\mathrm{b}$}  
\newcommand{\Atwo}{$A_2$}  
\newcommand{\Atworb}{$A_{2\mathrm{b}}$}  
\newcommand{\Aone}{$A_{\rm 1,z}$} 
\newcommand{\Jd}{$J_{\rm d}$}  
\newcommand{\Jh}{$J_{\rm h}$}  
\newcommand{\dJ}{$\dot{J}$} 
\newcommand{\dJd}{$\dot{J}_{\rm d}$} 
\newcommand{\dJh}{$\dot{J}_{\rm h}$} 
\newcommand{\Jdin}{$J_{\mathrm{d,in}}$}  
\newcommand{\Jdout}{$J_{\mathrm{d,out}}$}  
\newcommand{\Jhin}{$J_{\mathrm{h,in}}$}  
\newcommand{\Jhout}{$J_{\mathrm{h,out}}$}  
\newcommand{\Rcr}{$R_\mathrm{cr}$}  
\newcommand{\Rbar}{$R_\mathrm{b}$}  
\def\gtorder{\mathrel{\raise.3ex\hbox{$>$}\mkern-14mu
    \lower0.6ex\hbox{$\sim$}}}
\def\ltorder{\mathrel{\raise.3ex\hbox{$<$}\mkern-14mu
    \lower0.6ex\hbox{$\sim$}}}

\title{Dark matter halos and evolution of bars in disk galaxies:\\
collisionless models revisited}

\author{
Jorge Villa-Vargas\altaffilmark{1},
Isaac Shlosman\altaffilmark{2,3}
and
Clayton Heller\altaffilmark{4}
}
\altaffiltext{1}{
Department of Physics and Astronomy,
University of Kentucky,
Lexington, KY 40506-0055,
USA
}
\altaffiltext{2}{
JILA,
University of Colorado,
Boulder, CO 80309,
USA
}
\altaffiltext{3}{
National Institute of Standards and Technology,
Boulder, CO 80305-3328,
USA
}
\altaffiltext{4}{
Department of Physics,
Georgia Southern University,
Statesboro, GA 30460,
USA
}

\begin{abstract}
We construct and evolve three one-parameter families and one two-parameter 
family of steady-state models of stellar disks embedded in live dark matter 
(DM) halos, in order study the dynamical and secular phases of bar evolution. 
These models are tested against those published in the literature
in order to extend them and to include the gaseous component in the follow up 
paper. Specifically, we are interested in the angular momentum, $J$, 
redistribution in the disk-halo system during these two evolutionary phases
without distinguishing between the resonant and non-resonant effects. We 
confirm  the previous results and quantify for the first time the dual role
that the DM halos play in the bar evolution: 
more centrally concentrated halos dilute the dynamical processes of the initial 
bar growth, such as the spontaneous bar instability and the vertical buckling 
instability, and slow down the $J$ transfer, while facilitating it 
in the secular phase. The rate of $J$ transfer in the disk and the halo is 
followed up in order to identify sites and times of peak 
activity in $J$ emission and absorption. Within the corotation radius, \Rcr, 
the disk $J$ remains nearly constant in time, as long as \Rcr{} stays within 
the disk --- a sign that the lost angular momentum to the outer disk and the 
halo is being compensated by an 
influx of fresh $J$ due to the outward motion of \Rcr. We demonstrate that 
this is feasible as long as the bar slowdown dominates the loss 
of $J$ inside \Rcr. Next, we find that in some models the bar pattern speed 
stalls for prolonged time periods, i.e., the bar exhibits a constant rate of 
tumbling when \Rcr{} is located outside the disk. This phenomenon appears 
concurrent with the near absence of $J$ transfer between the disk and the halo, 
and is associated with the halo {\it emitting} $J$ at the corotation resonance 
and absorbing it at the inner Lindblad resonance. Furthermore, we confirm that 
stellar bars generally display the corotation-to-bar size ratios in the range 
of $\sim 1-1.4$, but only between the times of the first buckling and
\Rcr{} leaving the disk. Hence, the corotation-to-disk size ratio emerges as 
an important dynamic discriminator between various stages of barred disk 
evolution. Finally, we 
analyze a number of correlations between the basic parameters of a barred disk 
and a halo, some already reported in the literature and some new. 
\end{abstract}

\keywords{galaxies: evolution -- galaxies: halos -- galaxies: kinematics and 
dynamics -- galaxies: spiral --- galaxies: structure -- stellar dynamics --- 
cosmology: dark matter}

\section{Introduction}


Bars are expected to have a profound effect on disk galaxy evolution
because they constitute a major departure from an axial symmetry and hence
facilitate angular momentum and mass redistribution. Despite
prolonged and focused investigation of bar formation and evolution many
theoretical and observational issues remain unresolved. Among these are
the origin of bars in the universe --- spontaneous or tidally induced,
evolution of bar fractions with redshifts, and theoretical predictions
of bar pattern speeds. The intricacies of the bar growth or decay with time 
are obscure --- how do bars capture orbits? Do all bars extend to their
corotation radii? Lastly, various aspects of nested bar systems are only now
beginning to be analyzed.

Because the underlying dynamics of bars is strongly nonlinear,
their numerical modeling has spearheaded the efforts to understand their
orbital structure, ability to channel the angular momentum to the outer disk
and dark matter (DM) halo, efficiency in triggering the radial gas inflows, 
and more. While we have gained some insight to all of these processes, 
much remains to be done and one expects additional effects to surface.

In this work we aim at deeper understanding of angular momentum 
redistribution in the disk-halo system facilitated by the stellar bars in 
the presence of the gas component. Here, we analyze a set of 
equilibrium {\it collisionless}
models which differ by a single parameter from each other. In the
associated paper, we add the gas component to the system without affecting the
mass distribution there (Villa-Vargas et al., Paper~II, 
in preparation). Our collisionless modeling is in a way complementary
to that of Athanassoula (2003). We reproduce some of the 
correlations between the bar, disk and halo properties discussed in the
above work and quantify additional correlations found. All these
will be compared to models with gas.

Importance of a galactic spheroidal component in bar evolution has
been gradually understood, from its supposedly stabilizing effect on the 
disk (Ostriker \& Peebles 1973; Efstathiou et al. 1982), to serving as 
disk angular momentum sink (e.g., Sellwood 1980; Weinberg 1985; Debattista 
\& Sellwood 2000; Athanassoula 2002), and appears to be full of 
controversies (e.g., Christodoulou, Shlosman \& Tohline 1995a,b; 
Athanassoula 2008). Study of the dominant role of resonance interactions 
between the bar and surrounding
orbits was pioneered by Lynden-Bell \& Kalnajs (1972) and applied
to bar and halo orbits by Tremaine \& Weinberg (1984), Athanassoula (2002), 
Martinez-Valpuesta, Shlosman 
\& Heller (2006), Ceverino \& Klypin (2007), Weinberg \& Katz (2007a,b)
and Dubinski, Berentzen \& Shlosman (2009). Toomre (1981) argued that
the bar growth can be damped by the introduction of an inner Lindblad
resonance (ILR), by cutting off the swing amplification mechanism for the 
$m=2$ modes. Hotter stellar disks are expected to be more stable as well 
(e.g., Athanassoula \& Sellwood 1986). Counter-intuitively, in the long 
run, a disk embedded in a more centrally concentrated halo develops a 
stronger and longer bar (Athanassoula \& Misiriotis 2002).  

A comprehensive analysis
by Athanassoula (2003) has shown that the efficiency of resonances in 
angular momentum transfer
depends on the DM and stellar dispersion velocities and 
the DM densities in the vicinity of the major resonances.
As the resonances sweep across the phase-space, about $20\%-30\%$
of DM particles are locked in the lower order resonances at any given
time (Dubinski et al. 2009).  Additional aspects of the bar evolution are 
known to be
influenced by the spheroidal components, such as the vertical buckling
instability in the bar (Berentzen et al. 2007).

In general, $J$ transfer in the disk-halo systems can depend on three
parameters, namely, the particle population near low level resonances and
the velocity dispersions in the disk and halo (e.g., Athanassoula 2003). 
In principle, this means dependence on the halo central mass
concentration and on radial velocity dispersions in the disk and the halo.
We, therefore, focus on how parameters of the DM halo distribution,
such as its mass, concentration, extent, and rms velocities affect the 
disk-halo evolution and the angular momentum transfer mediated by 
the stellar bar. Specifically, we are interested in
how the presence of gas in the disk influence the angular momentum
redistribution in the system. For the purpose of clarity, we separate our
discussion of collisionless models (this Paper I) from models with gas (Paper II).
Collisionless models are used in order to verify the basic details of bar
evolution in order to serve as benchmarks for models with gas. They also
are used to calibrate our results to those of Athanassoula (2003).

Our paper is structured as follows: \S2 describes the numerics, initial
conditions and model parameters. \S3 provides results on basic evolution
of stellar bars in our sets of models and analyzes the redistribution of
angular momentum in these systems, while \S4 is focused on correlations
between various parameters in the bar-disk-halo models.
 
\section{Numerics and Modeling}


We use the $N$-body part of the FTM-4.4 hybrid code (e.g., Heller \& Shlosman
1994; Heller, Shlosman \& Athanassoula 2007b) to evolve the stellar disks and 
DM halos. The gravitational forces are calculated using the FalcON routine 
(Dehnen 2002) which scales as $O(N)$. 
The units of mass and distance are taken as $10^{11} \mathrm{M_{\odot}}$ and
10~kpc respectively. This makes the unit of time as $4.7 \times 10^7~\mathrm{yr}$
when $G=1$, and the velocity unit $208~\mathrm{km~s^{-1}}$.
The gravitational softening is $\epsilon_{\rm grav}=0.016$ for stars and DM
particles. The models consist of a stellar disk with  $N_{\rm d} = 2\times 10^5$ 
and of DM halo with $N_{\rm h} = 10^6$ particles.

\subsection{Initial conditions}
\label{s_inicon}

The initial conditions were created following
the prescriptions and density profiles from Hernquist (1990).
The mass volume density distribution in the disk is given in cylindrical 
coordinates by

\begin{equation}
\label{eq:rho_disk}
\rho_\mathrm{d} (R,z) = \frac{M_\mathrm{d}}{4\pi h^2 z_0} 
      \exp (-R/h)\ \mathrm{sech}^2 \left(\frac{z}{z_0}\right),
\end{equation}
where $M_\mathrm{d}$ is the disk mass, $h$ is a radial scale length 
and $z_0$ is a vertical scaleheight. 

The density of the spherical halo is given by
\begin{equation}
\label{eq:rho_halo}
\rho_\mathrm{h}(r) = \frac{M_\mathrm{h}}{2\pi^{3/2}} \frac{\alpha}{r_\mathrm{c}} 
                        \frac{\exp(-r^2/r_c^2)}{r^2+\gamma^2},
\end{equation}
where $M_\mathrm{h}$ is the mass of the halo, $r_\mathrm{c}$ is a Gaussian
cutoff radius and $\gamma$ is the core radius. $\alpha$ is the normalization 
constant defined by
\begin{equation}
\alpha = \{ 1 - \sqrt{\pi} q \exp(q^2) [1-\mathrm{erf}(q)] \} ^{-1}
\end{equation}
with $q=\gamma/r_\mathrm{c}$. 

The particle velocities, dispersion velocities and asymmetric drift
corrections have been calculated using moments of the collisionless Boltzmann
equation. Since models thus constructed are not in exact virial equilibrium, the 
halo component was relaxed for $t\sim 40$ in the frozen disk potential.

\subsection{Model parameters}

Our goal is to investigate the effect of the DM mass distribution on the
evolution of stellar bars in a live disk--halo system. We choose
a specific model, defined here as the standard model (SD), where the DM density 
profiles are specified by three parameters,
$M_\mathrm{h}$, $\gamma$ and $r_\mathrm{c}$, corresponding to the DM halo mass,
DM core radius and the DM Gaussian cutoff radius.
Based on the SD model, three sequences of models have been created, each 
sequence resulting from varying one of the parameters only.
In the fourth sequence, we simultaneously modify two parameters at the time, 
$\gamma$ and $r_\mathrm{c}$, in attempt to target the outer halo only, while 
trying to keep the inner halo unchanged.  
In the SD model, the halo-to-disk mass ratio in the inner $R=0.6$ is kept to 
unity. Table~\ref{sd_param} lists the values of the SD parameters.

\begin{table}[!h]
\centering
\caption{Parameters of the standard model}
\begin{tabular}{l r || l r}
\\
\hline
\multicolumn{2}{c||}{HALO} & \multicolumn{2}{c}{DISK} \\
Parameter & Value & Parameter & Value\\
\hline
$N_{\rm DM}$       &  $10^6$  &  $N_*$              &  $2\times 10^5$ \\
$M_\mathrm{h}$     &  3.15    &  $M_\mathrm{d}$     &  0.63 \\
$r_{\rm t}$        &  8.55    &  $R_{\rm t}$        &  1.71 \\
$\gamma$           &  0.1425  &  $h$                &  0.285 \\
$r_\mathrm{c}$     &  2.85    &  $z_\mathrm{0}$     &  0.057 \\
                   &          &  $Q$                &  1.5 \\
\hline
\end{tabular}
\tablecomments{$Q$ is the Toomre parameter fixed at $R=2.4h$, where $h$ is the
thickness of the disk; $r_{\rm t}$ and $R_{\rm t}$ are 
numerical truncation radii in the halo and the disk. All values are given in 
dimensionless units, \S2.}
\label{sd_param}
\end{table}

The initial mass volume density profile in the disk is kept unchanged in all 
models. The spatial distribution of the stellar particles is thus 
identical in all realizations of the initial conditions.
On the other hand, the velocities of the stellar particles have been adjusted in 
each models to provide for the changing rotational support against the combined 
gravitational potentials of the varying disk-halo systems. 
The four model sequences are as following:

\begin{description}
\item[$M_\mathrm{h}$ sequence:] the mass of the halo, 
     $M_\mathrm{h}$, has been reduced to $70\%$ and $40\%$ 
     (models M70 and M40 respectively) of the SD model.
\item[$\gamma$ sequence:] the DM core radius $\gamma$ was 
     increased by a factor 
     of 2 and 4 with respect to the SD model (models C30 and C57 respectively). 
\item[$r_\mathrm{c}$ sequence:] the DM Gaussian cuttoff
     radius $r_\mathrm{c}$ was decreased to $77\%$ and $56\%$ 
     of that in the SD model (models T22 and T16 respectively).
\item[$M_\mathrm{h}-r_\mathrm{c}$ sequence:] in these hybrid models, MT1
     and MT2, both $r_\mathrm{c}$ 
     and $M_\mathrm{h}$ have been varied, to keep the inner $r<0.1$ DM density 
     profile close to that of the SD model.
\end{description}
 
\begin{table}[!h]
\centering
\caption{Halo parameters for the model sequences}
\begin{tabular}{l r r r}
\\
\hline
Model     & $m_\mathrm{h}$ & $\gamma$ & $r_\mathrm{c}$ \\
\hline
SD        &  3.15  &  0.1425  &  2.85 \\
\\
M70      &  2.20  &  0.1425  &  2.85 \\
M40      &  1.26  &  0.1425  &  2.85 \\
\\
C30      &  3.15  &  0.30    &  2.85 \\
C57      &  3.15  &  0.57    &  2.85 \\
\\
T22      &  3.15  &  0.1425  &  2.20 \\
T16      &  3.15  &  0.1425  &  1.60 \\
\\
MT1      &  2.40  &  0.1425  &  2.20 \\
MT2      &  1.71  &  0.1425  &  1.60 \\
\hline
\end{tabular}
\tablecomments{All values are given in dimensionless units, \S2}
\label{seqpar}
\end{table}

\begin{figure}
   \centering
   \includegraphics[angle=-90,bb=59 149 558 394,width=0.8\linewidth]{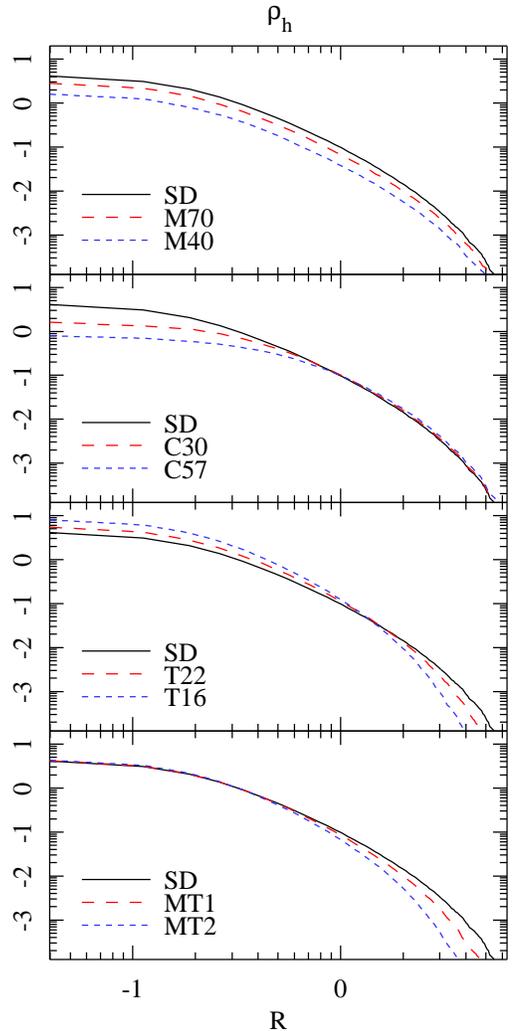}
   \caption{The DM halo density profiles at $t=0$ after relaxation in the
            frozen disk potential.
            Each panel shows the model sequences. From top to bottom: 
            $M_{\rm h}$, $\gamma$, $r_{\rm c}$ and the hybrid 
            $M_\mathrm{h}-r_\mathrm{c}$ sequences.  
             }
   \label{rhomods}
\end{figure}
\begin{figure}[!t]
   \centering
   \includegraphics[width=1.0\linewidth,bb=58 300 370 675]{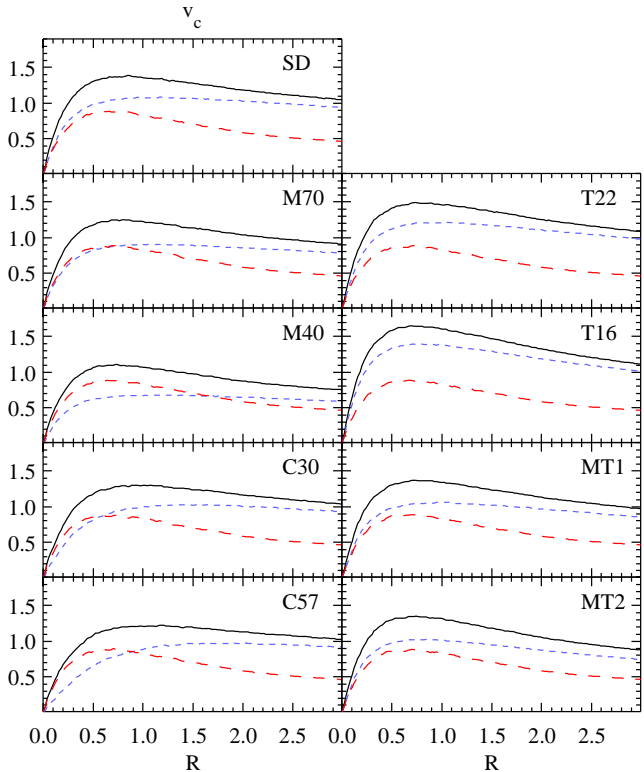}
   \caption{Circular velocities of all models at $t=0$ after halo relaxation
         in the frozen disk potential. Each panel displays the stellar
         disk rotation curve (red, long dashed), halo (blue, dashed) and
         the total curve (black, solid).
             }
   \label{vcir3}
\end{figure}

The DM halo parameters in each model are listed in Table~\ref{seqpar}.
The number of particles and the numerical truncation radii are the same as in 
the SD model. The halo mass $M_\mathrm{h}$ is conserved in sequences $\gamma$ and
$r_{\rm c}$. The DM density profiles of all models after relaxation in the frozen
disk potential are shown in 
Figure~\ref{rhomods}. The resulting DM density profiles exhibit monotonic decrease
along the $M_{\rm h}$ sequence, show a progressively larger core in the $\gamma$
sequence, move the outer halo mass inward to the inner halo, across $r\sim 1.4$,
in the $r_{\rm c}$ sequence,
and increase the outer halo mass while leaving the core unchanged in the hybrid
models sequence. The initial circular velocity curves, showing the contributions
from the disk, the halo and the total, are displayed in Figure~\ref{vcir3}.
Models M40, C30 and C57 host maximal disks which dominate the potential
of the inner part. Models SD, M70, MT1 and MT2 have equal contributions of
disk and halo at $t=0$, while T22 and T16 are halo dominated. With the
development of the stellar bar, the disk typically becomes even more dominating,
although the inner halo is also dragged in an adiabatic contraction
(e.g., Dubinski et al. 2009). 
 
\section{Results}

\begin{figure*}
\centering
   \includegraphics[angle=0,height=10.7cm,bb=111 49 471 716]{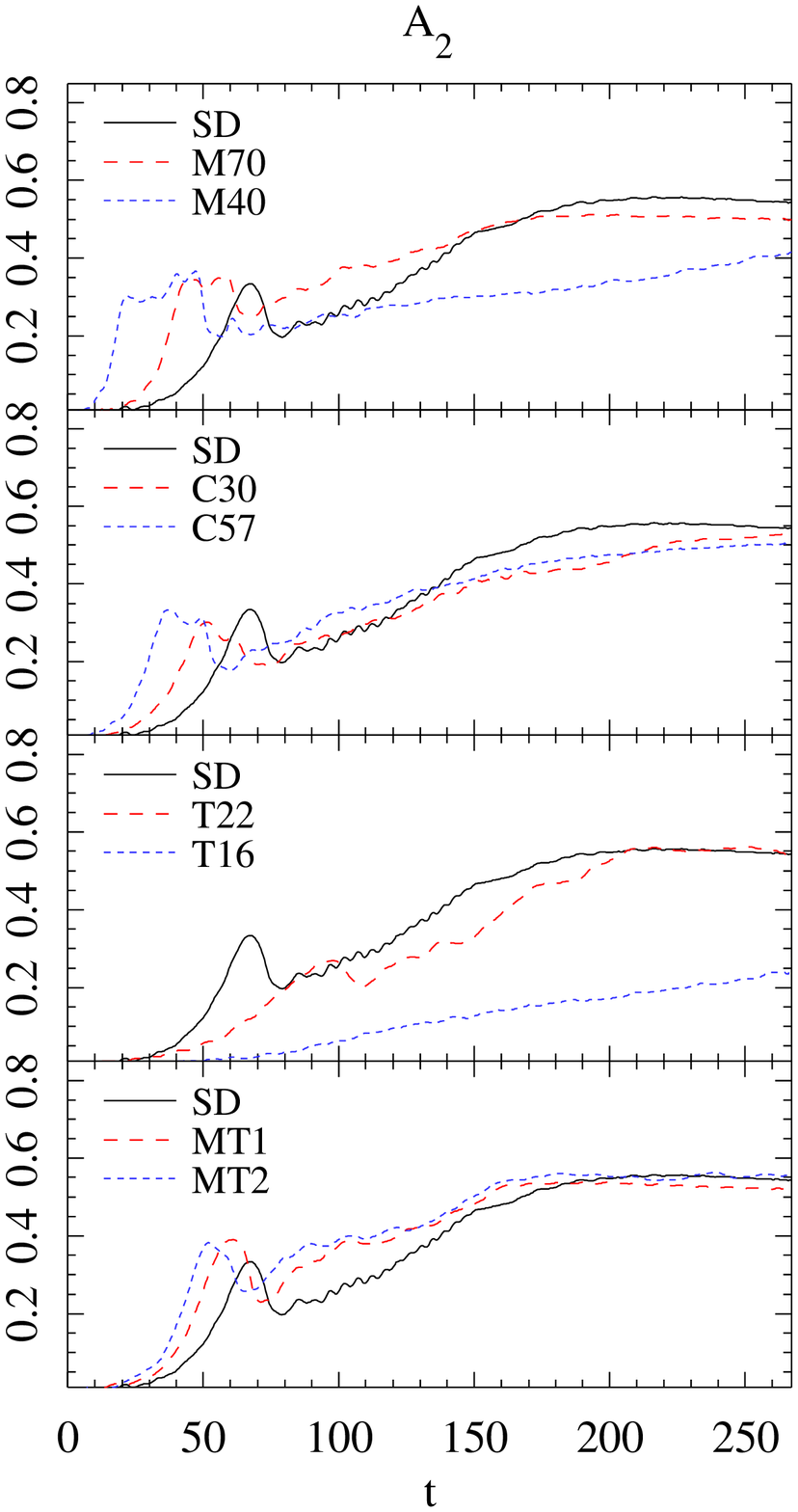}%
   \hspace{-1.15cm}\includegraphics[angle=0,height=10.7cm,bb=80 49 471 716]{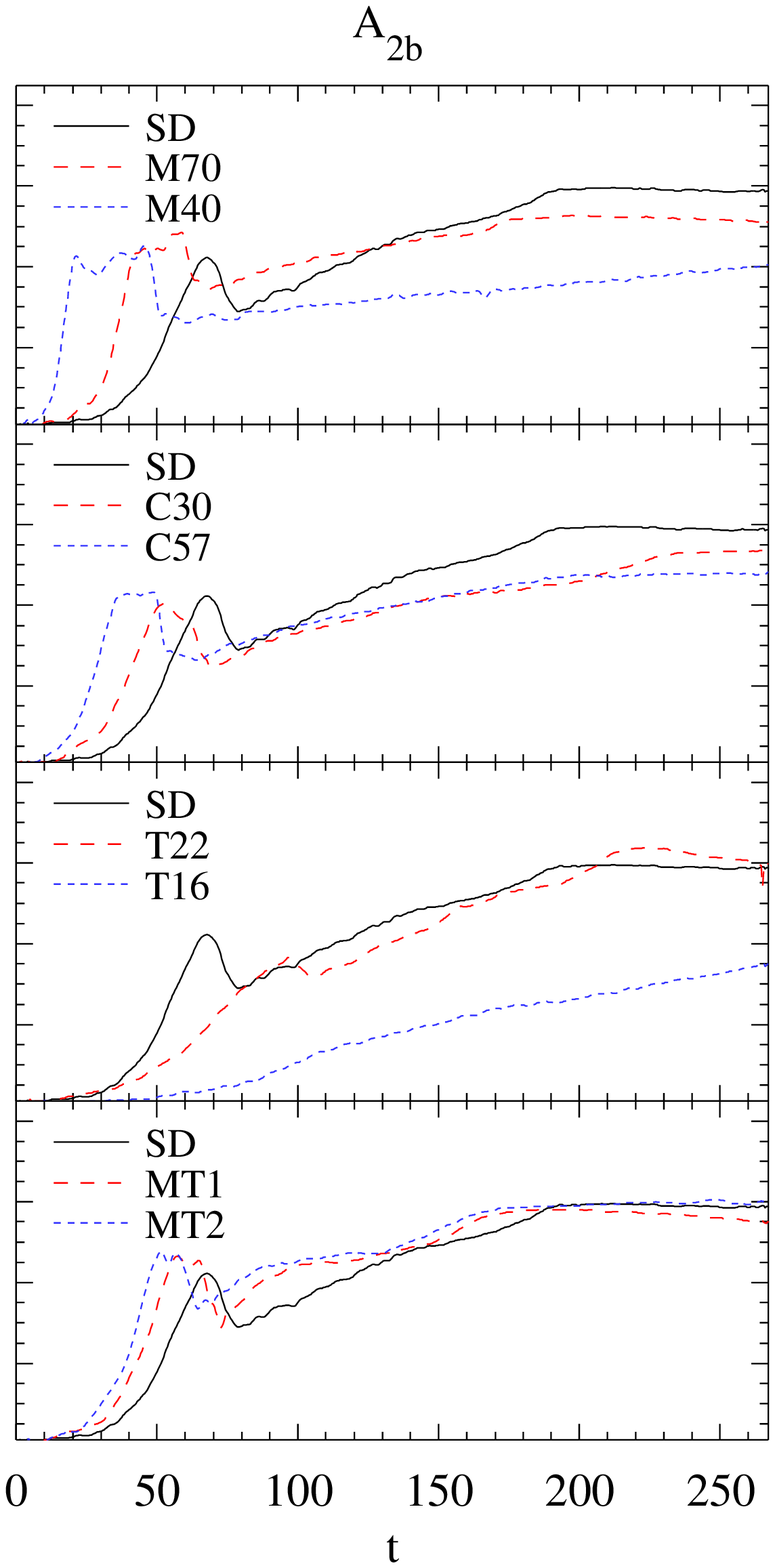}%
   \hfill\includegraphics[angle=0,height=10.7cm,bb=111 49 471 716]{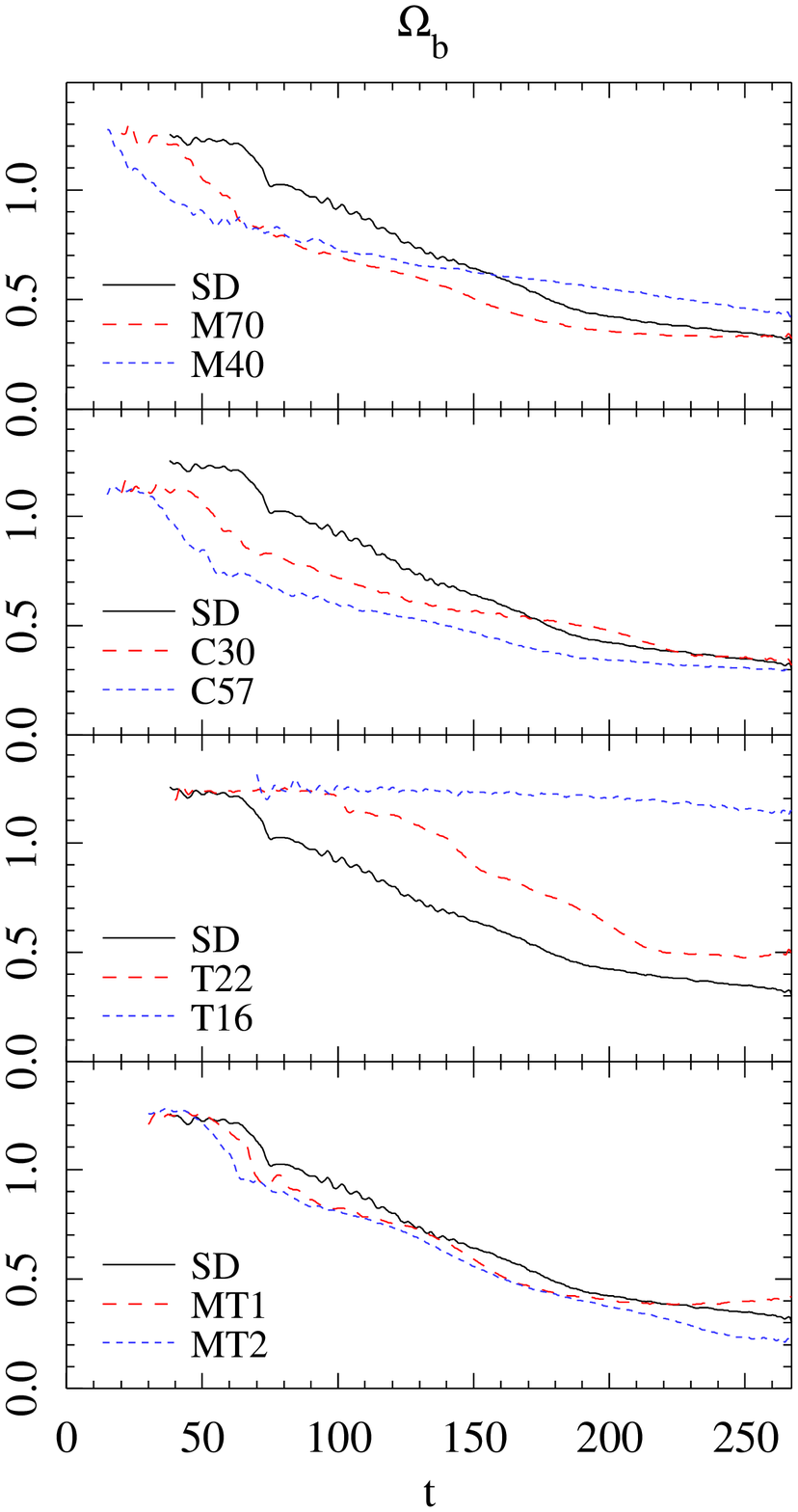}
   \caption{Time evolution of normalized bar amplitudes, \Atwo{}
        (left column) and \Atworb{}
        (middle column), and the bar pattern speed $\Omega_\mathrm{b}$ 
        (right column) in all models. The model sequences are indicated
        in the upper left corners.  
        All the data has been smoothed with a high frequency Fourier filter.}
   \label{Rb_A2}
\end{figure*}

The model sequences described in Table~2 differ only by the parameters of
DM distribution as given by Eq.~2. These include the total mass of the halo, its
core size and the Gaussian truncation radii. As the bar properties and angular 
momentum
evolution are both heavily dependent on the mass distribution in the system,
the choice of the free parameters allows us to fine-tune the changes. Because the DM
appears to be on the receptive side of the angular momentum transfer, we target
the halo as a whole, and its inner and outer parts separately. Discussion and 
comparison with published models is made in \S4.

\subsection{Bar strength and pattern speed evolution}

To gauge the strength of the bar we use the Fourier amplitude \Atwo{} of
the $m=2$ mode normalized by the $m=0$ mode.
It is obtained by various methods --- here we show the results of integration 
over all the disk, and integration over restricted cylindrical volumes of a 
particular interest. We define \Atwo{} in the integration limit over the fixed 
radial range $R = 0.1 - 1.8$. Second, we define  \Atworb{} when integrate
over $R = 0.1 - R_\mathrm{b}$ range, where \Rbar{} is the bar size 
defined in \S3.2. Intuitively, the second definition reflects the
bar properties more fully because \Atworb{} is not diluted by the disk
properties.
Both \Atwo{} and \Atworb{} show some similarities, e.g., the same peaks, 
raises and drops, and differences, e.g., the relative strength of the 
peaks (Fig.~\ref{Rb_A2}, first and second columns).

Most models show common stages in the evolution of \Atwo{} 
and \Atworb:
(1) an initial exponential steep rise, (2) a subsequent peak or plateau 
followed by a sudden drop, ensued by (3) a more gradual and sustained
rise, ending by (4) a saturation of the bar strength. 
The duration of each stage varies from model to model, but stages ``1'' and 
``2'' are much shorter than stage ``3''. We refer to stages ``1'' and 
``2'' as the dynamical evolution of the bar, and stages ``3'' and ``4'' as the 
secular evolution of the bar.
Models M40 and T16 are the exceptions. The former model has the least 
massive halo and exhibits a prolonged 
secular growth of the bar after the first peak. 
The bar amplitude does not reach saturation even in the Hubble time.  
The latter model has the most massive halo core and the bar strength never 
shows a clear
first peak seen in all other models. However a careful analysis of \Atwo{}
and especially \Atworb{} reveals that the exponential growth is terminated
at $t\sim 120$ and the subsequent evolution lucks the vertical buckling
instability (see \S3.4), but otherwise follows the path of other models.
We take a closer look at these trends 
in \S4.  We note that our decision to create one-parameter sequences
naturally leads to a wide spectrum of bar properties and can include some
of the extreme behavior.

Some clear trends of the bar strength behavior can be observed along each 
of the sequences defined in
\S2.2. First, in all models, except the two exclusions mentioned above, 
\Atwo{}\ and \Atworb{}\ have saturated by $t\sim 230$. Because \Atworb{}\
is tailored for the bar, the saturation of this parameter is more obvious
(more about this in \S3.3). Second,
making the {\it inner} halo less massive (i.e., less centrally concentrated), 
either by reducing the total
halo mass or by increasing the size of the central core (sequences $M_{\rm h}$ 
and $\gamma$, respectively), results in a shorter rise time of the bar 
instability and hence brings up the bar earlier. Third, increase in the
mass of the inner halo and decrease in the outer one, has the most dramatic
effect on the bar strength, substantially increasing the timescale of
the bar instability. This hints at DM mass concentration being important
rather than the total halo mass. Varying the mass of the outer halo alone 
has a much smaller effect on the bar.
Finally, in all one-parameter sequences, the first \Atwo{} or \Atworb{} peak 
forms a progressively more extended plateau for less concentrated halos,
respectively. Again T16 is an exclusion. 

\begin{figure}[ht!!!!!]
\centering
   \includegraphics[angle=0,scale=0.55]{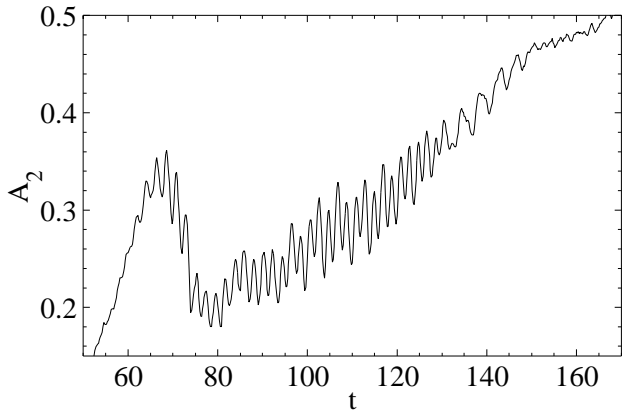}%
   \caption{Close up of the \Atwo{} evolution in the SD model, showing the 
        high frequency oscillations during $t\sim 70-160$.  
        No smoothing has been applied to the data.}
   \label{A2SD}
\end{figure}

Furthermore, \Atwo{} and \Atworb{} show high frequency variability in some
models (e.g., Figure~\ref{A2SD}), being strongest in models SD, C30 and T22.
This variability can also be seen in \Rbar{} and \ps{} evolution, the bar
length and its pattern speed. This variability is limited to the time of
the bar secular growth and dies out with saturation of the bar strength,
\Atwo{}\ and \Atworb{}. These oscillations coincide in time with the
presence of four short arms located close to the bar end, and two spirals
in the outer disk beyond the bar radius. Both sets of spirals have pattern
speeds larger than that of the bar.
The presence of more than one strong pattern of waves or modes in the disk
can be the result of a non-linear mode coupling (e.g., Tagger et al. 1987;
Sygnet et al. 1988) that gives rise to beat waves of a frequency
$\omega_{\rm beat} = \omega_1 \pm \omega_2$ and the azimuthal wave number 
$m_{\rm beat} = m_1 \pm m_2$,
where $m_1$ and $m_2$ are the (azimuthal) wave numbers of the interacting waves
and $\omega_1$ and $\omega_2$ are their frequencies, $\omega=m\Omega$, where
$\Omega$'s are their pattern speeds.
This type of coupling has been observed and quantified in numerical 
simulations (e.g.,
Masset \& Tagger 1997; Martinez-Valpuesta 2006). Debattista \& Sellwood (2000)
has also observed the high frequency variability, but no quantitative 
analysis of the mode coupling was attempted.

We test whether the bar and the outer two spirals, both $m=2$ modes, 
couple to give rise to the observed beat mode $m_{\rm beat}=4$ arms.
We verify that the frequency $\omega_{\rm beat}$ of the $m_{\rm beat}=4$ mode 
is actually the sum of the
frequencies of the interacting modes. Table~\ref{m2m4spectra} shows the
frequencies $\omega_{\rm b}=2$\ps, $\omega_{\rm s}=2\Omega_{\rm s}$ and 
$\omega_{\rm beat}$ obtained with the mode coupling
analysis, with an error of $\pm 0.02$. Here $\Omega_{\rm s}$ is the pattern
speed of the outer $m=2$ spirals.
The measured frequency of the
beat mode (column 5) matches the frequency predicted by the non-linear
coupling (column 6). We can explain the oscillations in \Atwo{} as resulting 
from superposition of the major axis of the bar with the outer spirals, which 
are the second strongest mode (after the bar itself) present in the disk.
The frequency $f_{\rm A2}$ of the \Atwo{} oscillations must thus be equal to 
the frequency at which the spirals align with the bar
$f_\mathrm{sb}\equiv (\omega_{\rm s}-\omega_{\rm b})/2\pi$. As can be seen 
from columns~7 and 8, these two frequencies coincide within the error 
limits.

\begin{table}[!h]
\centering
\caption{Mode coupling in the disk}
\begin{tabular}{l *{7}{r}}
\\
\hline
Model & Time & $\omega_{\rm b}$ & $\omega_{\rm s}$ & $\omega_{\rm beat}$ & $\omega_{\rm b} + \omega_{\rm s}$
     & $f_{\rm A2}$ & $f_\mathrm{\rm sb}$ \\
\hline
SD   & 106 & 1.80 & 4.86 & 6.66 & 6.66 & 0.49 & 0.49 \\
C30  & 113 & 1.32 & 4.24 & 5.58 & 5.56 & 0.45 & 0.46 \\
T22  & 149 & 1.80 & 5.72 & 7.58 & 7.52 & 0.61 & 0.62 \\
\hline
\end{tabular}
\tablecomments{(1) the model; (2) the time at which the measurements were taken;
              (3) $\omega_{\rm b}$; (4) $\omega_{\rm s}$; (5) $\omega_{\rm beat}$;
              (6) the expected frequency of the beat mode $\omega_{\rm b} + \omega_{\rm s}$;
              (7) measured frequency of \Atwo{} oscillations
              (8) frequency of alignment of bar and spirals.}
\label{m2m4spectra}
\end{table}

Figure~\ref{Rb_A2} shows the evolution of \ps{}  in our models, which
is well defined by the time \Atwo{}~$\sim 0.05$. In all models, the 
initial value of \ps{} is very similar, with the scatter of $\pm 0.1$.
Although the details of the evolution of \ps{} differ from model to model, 
most importantly, often shows an (anti)-correlation with \Atwo{}
and \Atworb, as emphasized already by Athanassoula (2003). This is not
always observed in cosmological simulations of disk evolution 
(Heller, Shlosman \& Athanassosula 2007a; Romano-Diaz et al. 2008). 
We find, however, that the anti-correlation between the bar strength and its 
pattern speed are limited to times prior to saturation of \Atwo{} and \Atworb.
In \S3.3 we show that this corresponds to times when the bar corotation
radius lies within the disk. In three models, M70, T22 and MT1, the \Atwo--\ps{}
correlation is maintained even after the corotation leaves the disk ---
in these models the total $J$ of the disk is conserved during this time
period of flat \Atwo{} and \ps{}.

A close examination of \ps{} evolution in Fig.~\ref{Rb_A2} reveals prolonged
periods of \ps$\sim const.$ These are found either in the dynamical phase,
prior to the first buckling but when the bar is already sufficiently strong 
(SD, M70, C57), or during the late stage of secular
evolution, especially pronounced in M70, T22 and MT1, and to a lesser degree in 
C57. These latter cases are of most interest to us.
A similar behavior was discussed before by Valenzuela \& Klypin (2003),
who related it to an abnormally low dynamical friction of the bar against
the background. We return to this issue in \S3.6.

\subsection{Evolution of the bar length}

\begin{figure*}[ht!!!!!!!!!!!!]
   \centering
   \includegraphics[width=0.25\linewidth,bb=100 17 492 734]{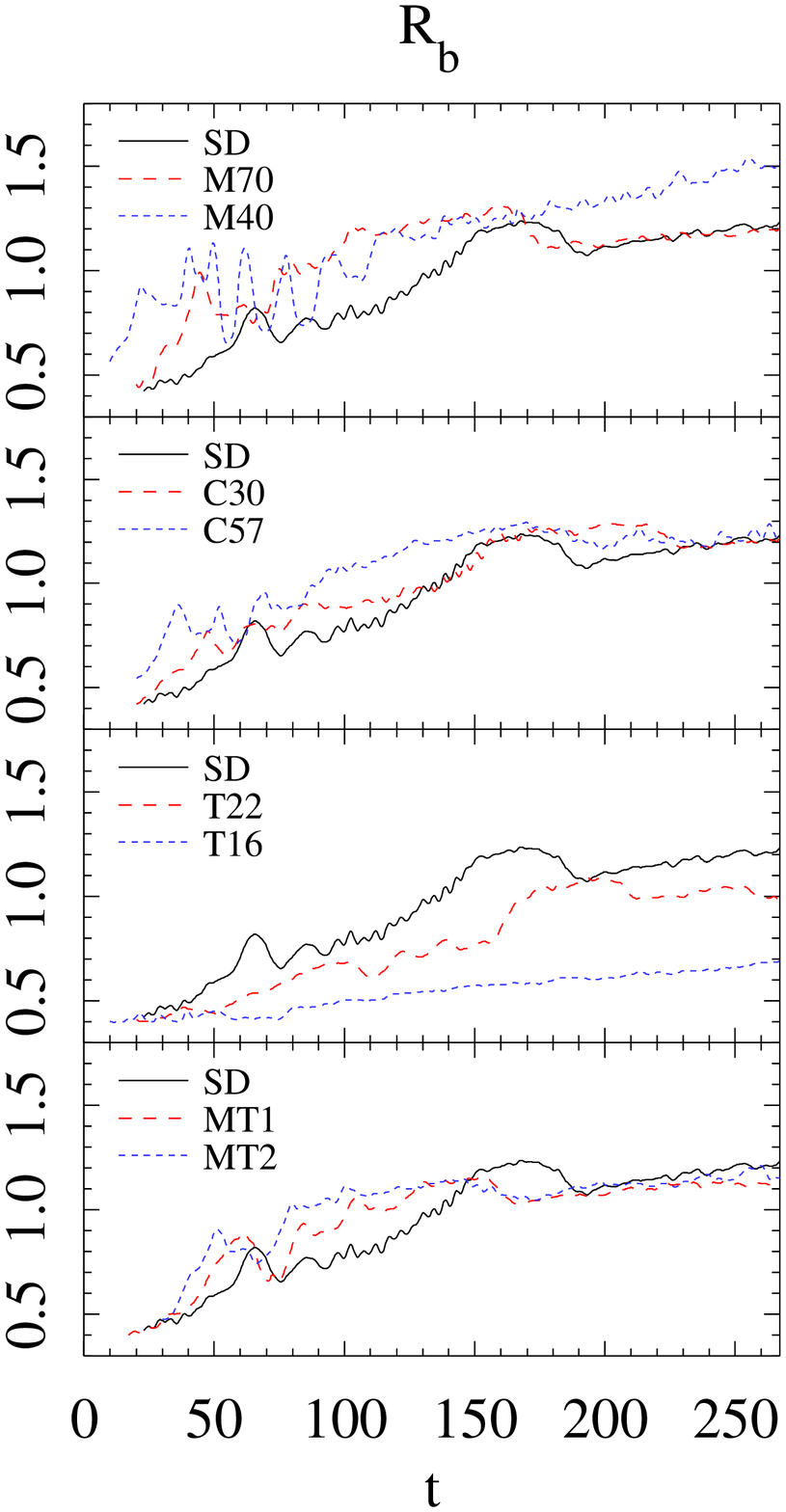}%
   \hfill\includegraphics[width=0.25\linewidth,bb=100 17 492 734]{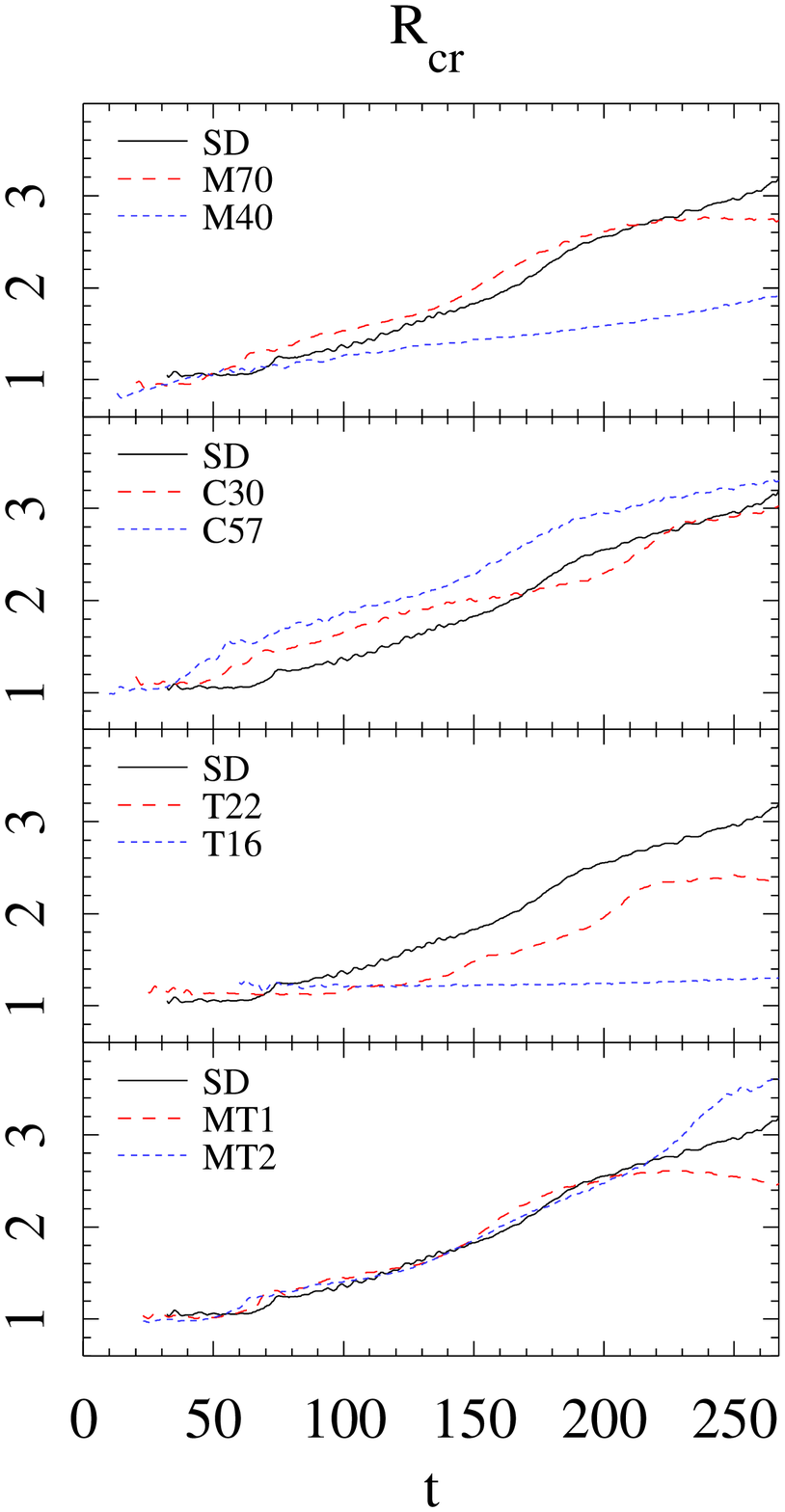}%
   \hfill\includegraphics[width=0.25\linewidth,bb=100 17 492 734]{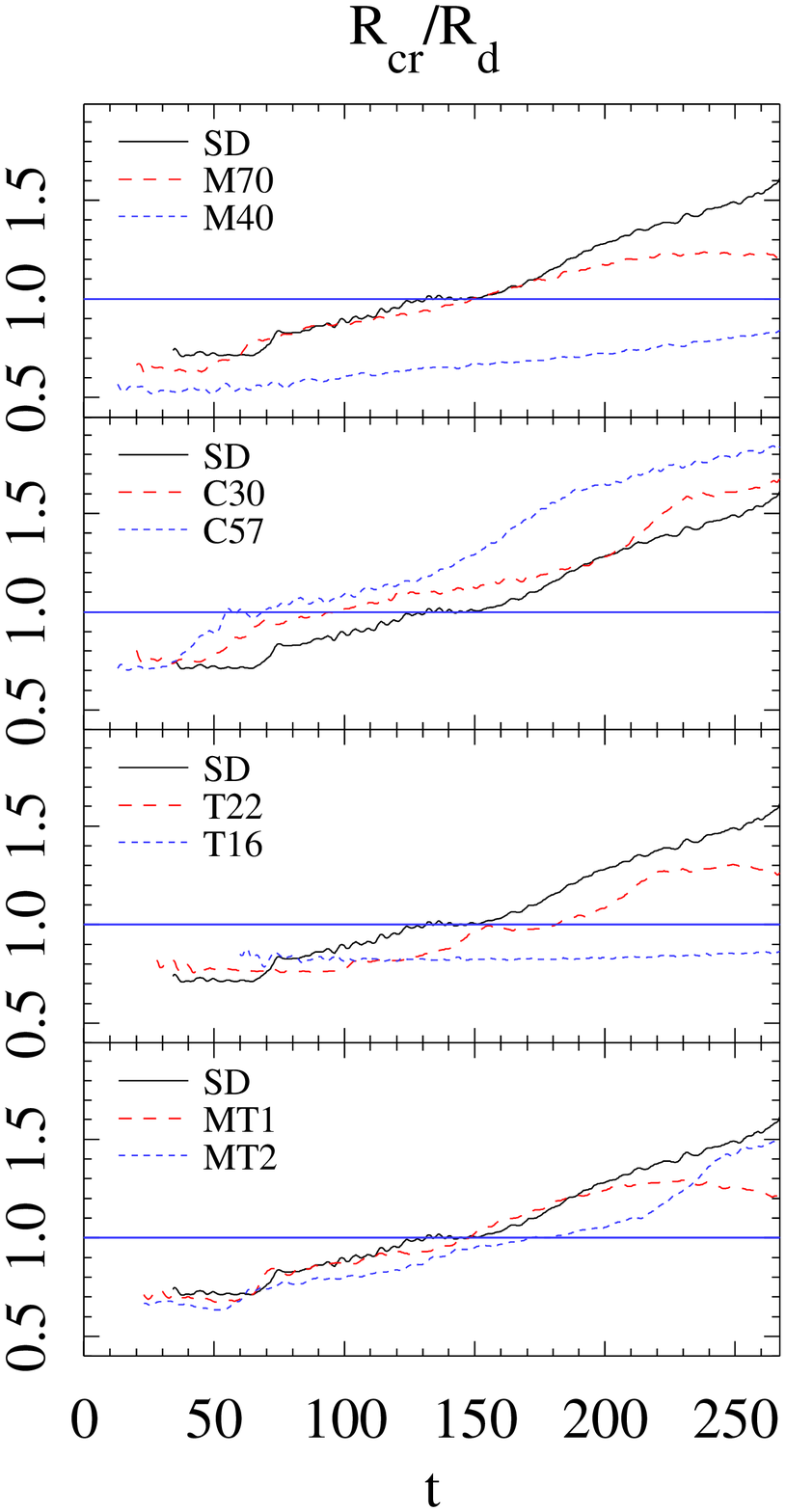}%
   \hfill\includegraphics[width=0.25\linewidth,bb=100 17 492 734]{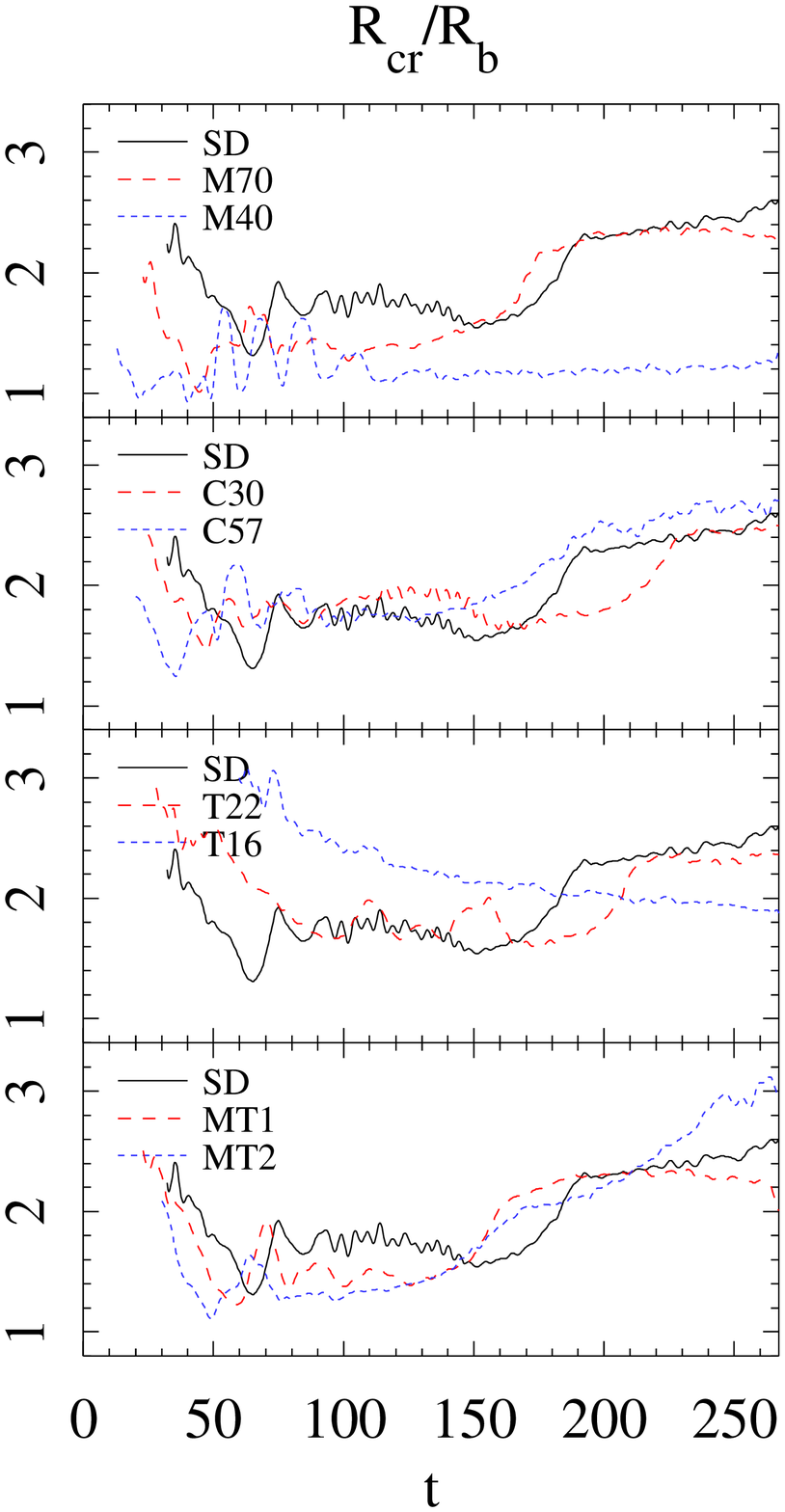}
   \caption{Time evolution of (from left to right) \Rbar,
        \Rcr, \Rcr/$R_{\rm d}$ --- ratio of \Rcr{} to the radius of
        the disk, and \Rcr/\Rbar.
        The \Rcr/$R_{\rm d}=1$ line has been drawn as a reference.
        SD model (continuous line) is shown in each panel for comparison.
}
   \label{Rcd}
\end{figure*}

The semi-major axis of the bar, \Rbar, is defined here as the radius 
where the bar equatorial ellipticity drops by $15\%$ off its peak. 
Martinez-Valpuesta et al. (2006) has tested this method by contrasting it
with the most reliable way of determining the bar size using the last stable 
orbit supporting the bar (Martinez-Valpuesta et al. 2006). It seems robust
when applied after the first maximum of the bar strength.
The ellipticity of the bar at different radii is obtained by fitting 
ellipses to the isodensity contours in the face-on disk. 
Figure~\ref{Rcd} displays the evolution of \Rbar{} for all the models.

Similarly to the evolution of \Atwo{} and \Atworb{}, the bar
length has initial period of a fast growth that reaches maximum, 
sometimes followed by a drop and subsequent sustained but 
slower rise, in most models. Some models show a degree of saturation, 
toward the end. The first peak in \Rbar{} always coincides in time with that 
of \Atwo{}, except in the model T16 which lacks it.
The peak in \Rbar, however, is not as pronounced as that in \Atwo{} and there
are additional caveats. As in \Atwo{}, the initial growth varies with a model 
in such a way that, in models with the less centrally concentrated halos, 
the time span of the initial growth is shorter. 
Furthermore, in some cases, additional drop in the bar length can be seen 
after some time
of the secular growth, in the time range of $t\sim 150 - 260$. This drop 
appears to be related to the formation and disappearance of the {\it ansae} at 
each tip of the bar (Martinez-Valpuesta et al. 2006). The ansae 
eventually `detach' 
from the bar and becomes misaligned with its major axis. The transition to 
this misalignment results in a fast decrease 
of the bar length, taking place at $t\sim 180$ for model SD, $\sim 190$ for 
M70, 22 for C30, 190 for C57, 205 for T22, and 165 for MT1 and MT2. 
 
In general, evolution in the bar length does not necessarily go in 
tandem with the changes in \Atwo{} and is much less monotonic. 
For example, the drop in \Rbar{} around $t\sim 180$ in the SD model 
has no counterpart in the \Atwo{} or \Atworb{} evolution.
Of course, a stronger \Atwo{} does not mean a longer bar, 
as seen in the last $\Delta t \sim 80$ of the evolution of the sequence 
$M_\mathrm{h}$ models. 

Overall, the $M_{\rm h}$ and $r_{\rm c}$ sequences show a substantial
dispersion in the bar sizes at the end of the simulations, while two other
sequences end up with very similar bars. In all sequences, the bars differ
substantially during the intermediate times, especially after the first 
buckling time.

\subsection{Evolution of the bar corotation radius}

The bar corotation radius \Rcr{} has been computed using linear approximation 
(the second column in Fig.~\ref{Rcd}). It grows with time as a consequence of 
the bar slowdown. \Rcr{} keeps growing throughout the 
simulation, the exceptions are models M70, T22 and MT1, in which a plateau is 
reached. We use this characteristic radius in order to construct and follow
the evolution of two important ratios, namely, \Rcr/\Rbar{} and \Rcr/$R_{\rm d}$.
 
The position of \Rcr{} has a profound effect on the angular momentum transfer
from the bar region to the outer disk and the DM halo (Athanassoula 2002; 
Martinez-Valpuesta et al. 2006). The DM particles at the corotation resonance 
are responsible for much of the angular momentum absorption by the halo. Hence, 
the availability of halo or
outer disk resonant particles is of the prime importance to the evolution of 
the bar. Specifically, it is important to know whether \Rcr{} 
lies within the stellar disk in our models at all times. 
We define $R_{\rm d}$ as the disk radius which encloses 98\% of the 
disk mass, and the evolution of the ratio  \Rcr/$R_{\rm d}$ is shown in 
Figure~\ref{Rcd}. At the time of the bar formation, \Rcr/$R_{\rm d}$ lies 
between 
0.5 and 0.8 in all models. This ratio raises with time enough to 
move the corotation out of the disk with the exception of models 
M40 and T16, where bars grow at the slowest pace during secular 
evolution. The growth of the bar seems to be sensitive to the 
moment at which \Rcr{} reaches the disk edge --- even if the growth continues,
it proceeds at a slower pace. This change happens with a slight but measurable 
delay of $\Delta t\sim 20$. Moreover, as noted before, 
the detachment of the ansae from the bar happens just before 
\Rcr/$R_{\rm d}\sim 1$, which manifests itself by a sharp drop in the bar length. 
 
We pay a special attention to the bar evolution when \Rcr/$R_{\rm d}\sim 1$.
Fig.~\ref{Rcd} displays an interesting behavior of this ratio in the
above regime --- it flattens with time before resuming its growth.
Analyzing this behavior, we find that as the \Rcr{} moves outside the disk, 
some stellar particles become trapped at \Rcr{}.
$m=2$ spiral arms appear connecting the outer edge of the disk
to these trapped particles. This is characteristic of all models where
the above conditions exist. At $t\sim 200-250$, depending on the model,
the spirals dissolve, and their particles
form an amorphous cloud outside the disk. We return to this issue in \S3.6.  

The bar must lie of course inside its \Rcr, because orbits outside \Rcr{}
do not support the bar. The morphology of the offset (gas) shocks
in the numerical bars has been argued to constrain the ratio
\Rcr/\Rbar, restricting it to \Rcr$/R_{\rm b}=1.2\pm 0.2$ (Athanassoula 1992;
see also early work by van Albada \& Sanders 1982). 
Limiting our discussion of this ratio only to times when \Rcr/$R_{\rm d}\leq 1$,
we observe it dropping to the range $\sim 1.0-1.4$ after the first buckling, and
staying generally below $\sim 1.8$ until the \Rcr{} moves out of the disk.
Model T16 (no buckling!) does not follow this trend and rather shows a 
monotonic decay to $\sim 1.9$ at the end of the run. On the other hand,
M40 --- the only other model with \Rcr{} always in the disk, displays 
a remarkably constant  \Rcr$/R_{\rm b}=1.2\pm 0.2$ almost over the entire
run (Figure~\ref{Rcd}). Naturally, after \Rcr{} moves out of the disk,
this ratio is bumped to above 2. In Paper~II, this ratio will be tested against
the shape of the offset shocks in the gas.

\begin{figure}[ht!!!!!!!!]
   \centering
   \includegraphics[width=0.8\linewidth,bb=100 17 492 734]{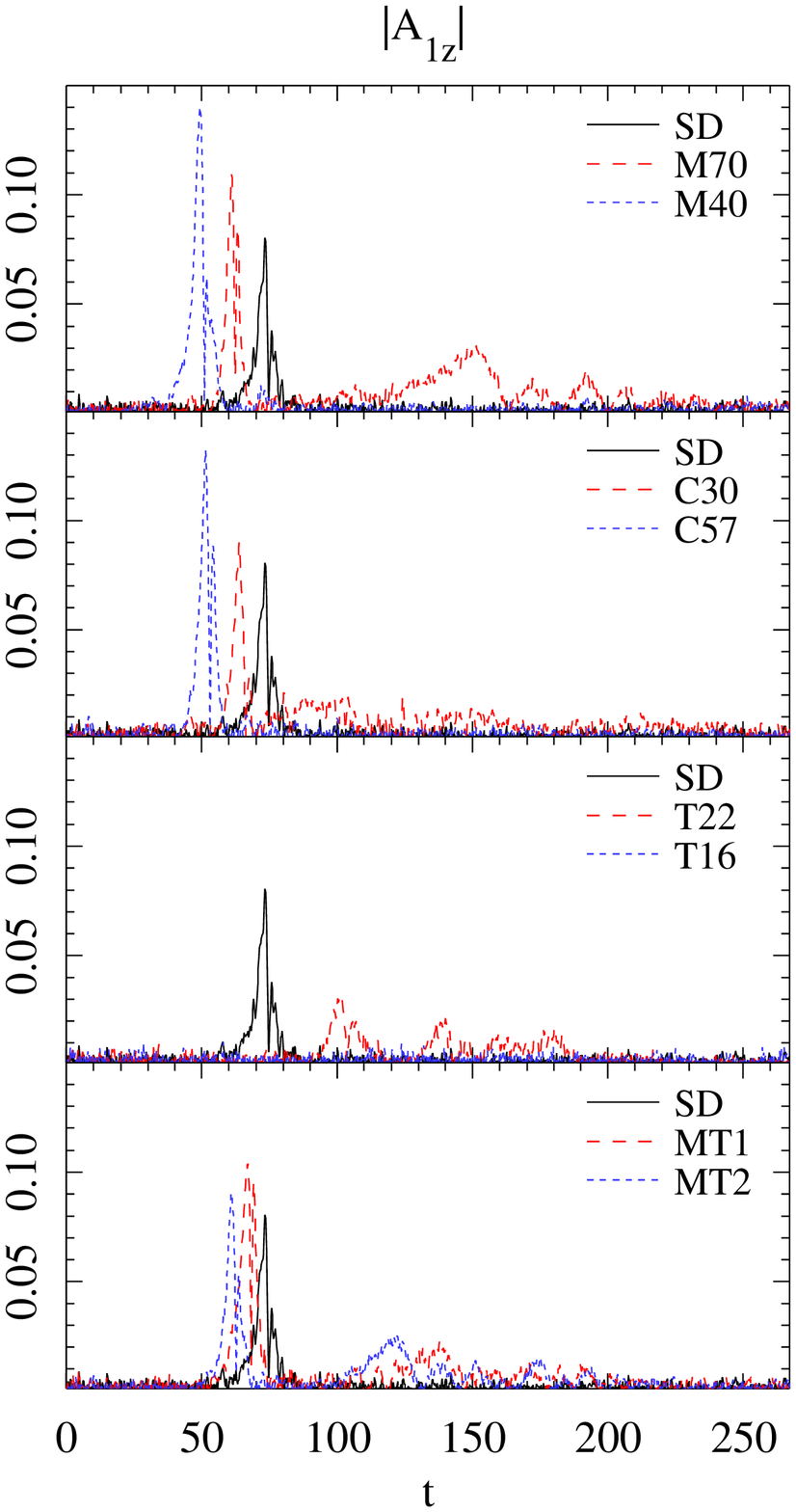}
   \caption{Evolution of the bar vertical asymmetry given by
   the Fourier coefficient $|A_{\rm 1,z}|$ of the $m=1$ mode in the $rz$-plane 
   corotating with the bar major axis. No filtering was applied to this data. 
   }
   \label{Rb}
\end{figure}

\subsection{Vertical buckling in the bar}

The $N$-body bars undergo phases of vertical buckling instability 
where transient asymmetries develop in the form of vertical oscillations 
of the equatorial plane and asymmetric thickening of the disk (e.g., Combes \&
Sanders 1981; Combes et al. 1990; Pfenniger \& Friedli 1991; Raha et al. 1991), 
and results in the formation of the so-called peanut/boxy-shaped bulges (Kormendy
\& Kennicutt 2004 and refs. therein).
These phenomena can be recurrent and affect different parts of the bar 
(Martinez-Valpuesta et al. 2006; Berentzen et al. 
2007). We measure the buckling asymmetry with the Fourier coefficient \Aone{}
of the $m=1$ component in the $rz$-plane, again normalized by the m=0 mode.
In all our models, except T16, we detect at least a single phase of 
the vertical buckling. 
The buckling always happens simultaneously with the first \Atwo{} drop (e.g.,
Martinez-Valpuesta \& Shlosman 2004), which we confirm here. 
In all the model sequences, models with less centrally concentrated 
halo exhibit earlier and stronger buckling. This happens because \Atwo{}\
reaches its first maximum earlier in these models. We do not find any clear
correlations between the strength and time of the second buckling and the 
initial properties of the models. In all cases of repeated bucklings, \Rcr{} 
lies within the disk.  

We note that T16 does not exhibit the drop in \Atwo{} or \Atworb{}, but does 
show a break in their slopes around $t\sim 120$. Moreover, while all our
models develop peanut/boxy bulges abruptly after these drops, T16 develops
this bulge gradually over the secular stage of its evolution. This model
appears as a nice example of a secular buildup of peanut/boxy bulges as
a result of the diffusive action of the vertical inner Lindblad resonance
(Friedli \& Pfenniger 1990; Martinez-Valpuesta \& Shlosman 2004). 
Berentzen et al. (2007) have shown that mass concentrations resulting
from gas accretion to the center damp the vertical buckling and alter
the bulge shape, making it more elliptical, if the gas fraction in the disk
is high. T16 is a purely collisionless 
model with the most centrally concentrated halo. It is possible that a
qualitatively different behavior is the corollary of this concentration.
Clearly,  the action of the vertical resonance leads to a {\it secular} and
not dynamical buildup of a peanut-shaped bulge in this case.

We also note that in all models the nonlinear inner Lindblad resonance
(ILR) first appears only after the first buckling, although the linear 
analysis claims their existence from the start (Pfenniger \& Friedli 1991; 
Martinez-Valpuesta et al. 2006). The appearance of this resonance is 
the consequence of the 
increase in the central mass concentration, by a factor of 
$\sim 2$, as a result of the bar buckling, which drags inward the DM as 
well (Dubinski et al. 2009).

\begin{figure}[!ht]
  \centering
\includegraphics[width=0.8\linewidth,bb=100 17 492 734]{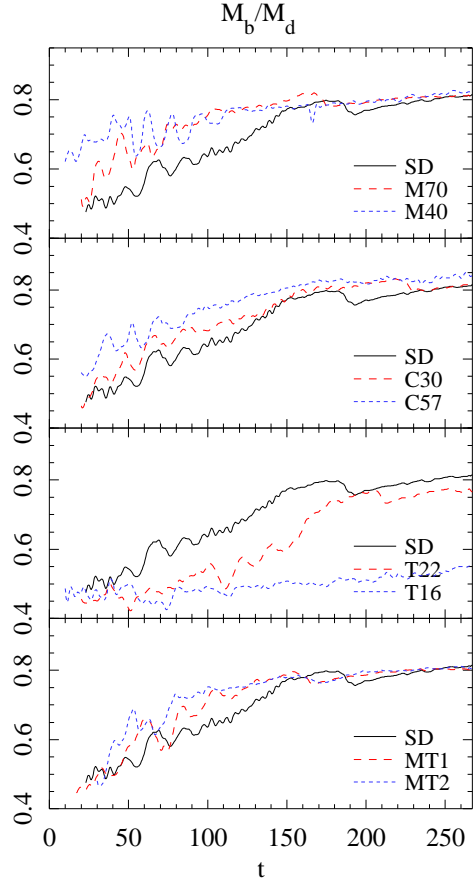}
  \caption{Evolution of bar-to-disk mass ratio $M_{\rm bar}/M_{\rm d}$.  
}
  \label{MbMd}
\end{figure}

\subsection{Evolution of bar-to-disk mass ratio}

The bar increases its mass by capturing particles from the disk within its
\Rcr. In Figure~\ref{MbMd} we plot the fraction of the stellar mass contained
inside the bar, $M_{\rm bar}/M_{\rm d}$. To estimate the bar, we sum 
the particle masses
within a rectangular box aligned with the bar major axis, with
dimensions given by the major and minor axes of the bar and $|z| \le 0.5$.

By the end of the run, the bar has captured between $75\%$ and $85\%$
of the disk mass (the exception is model T16
whose bar growth has been severely delayed by the halo mass concentration).
The initial stage of bar formation is associated with an intense capture
of disk particles. Eventually this rate
declines, and even though the bar mass keeps growing it does so at a much
more modest pace. The growth of $M_{\rm bar}/M_{\rm d}$ clearly
correlates with the evolution of the bar size in Fig.~\ref{Rcd}
(left frame) --- an increase/decrease in one of them corresponds with an
increase/decrease in the other. Typically, the growth of the bar size and of 
$M_{\rm bar}/M_{\rm d}$ saturates when the \Rcr/$R_{\rm d}$ is driven 
above unity. However, surprisingly, there are exclusions of this behavior, 
e.g., C57.
  
We do not find correlation between the bar mass
and \Atwo. It may happen that the bar keeps capturing mass while
\Atwo\ is saturated, as in model SD at $t\sim 210$ and later on.
In summary, the ratio  $M_{\rm bar}/M_{\rm d}$ saturates around 0.8 in
in all models, except T16.  

\begin{figure*}
   \centering
   \includegraphics[width=0.3\linewidth,bb=115 41 463 716]{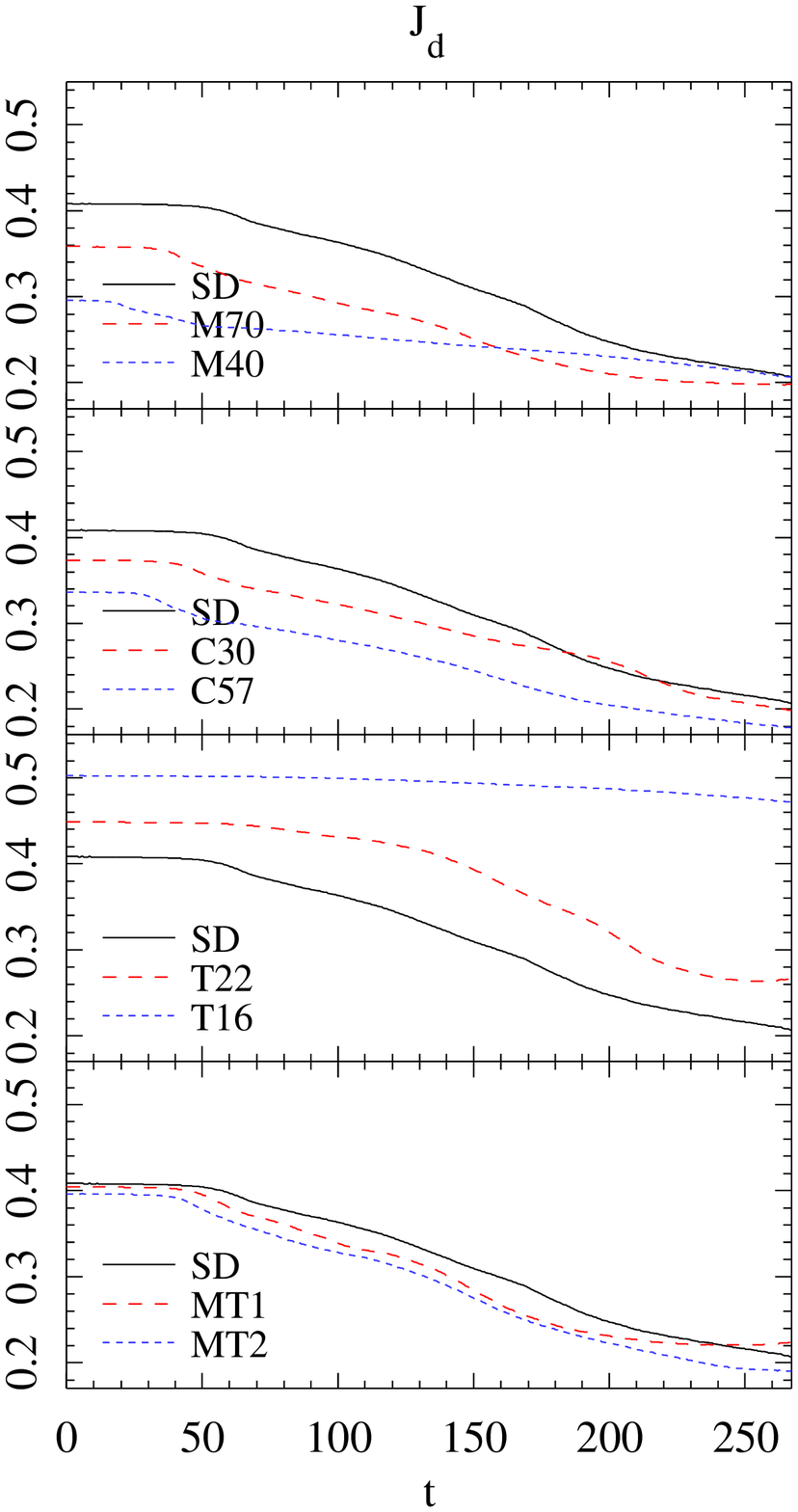}%
   \hfill\includegraphics[width=0.3\linewidth,bb=115 41 463 716]{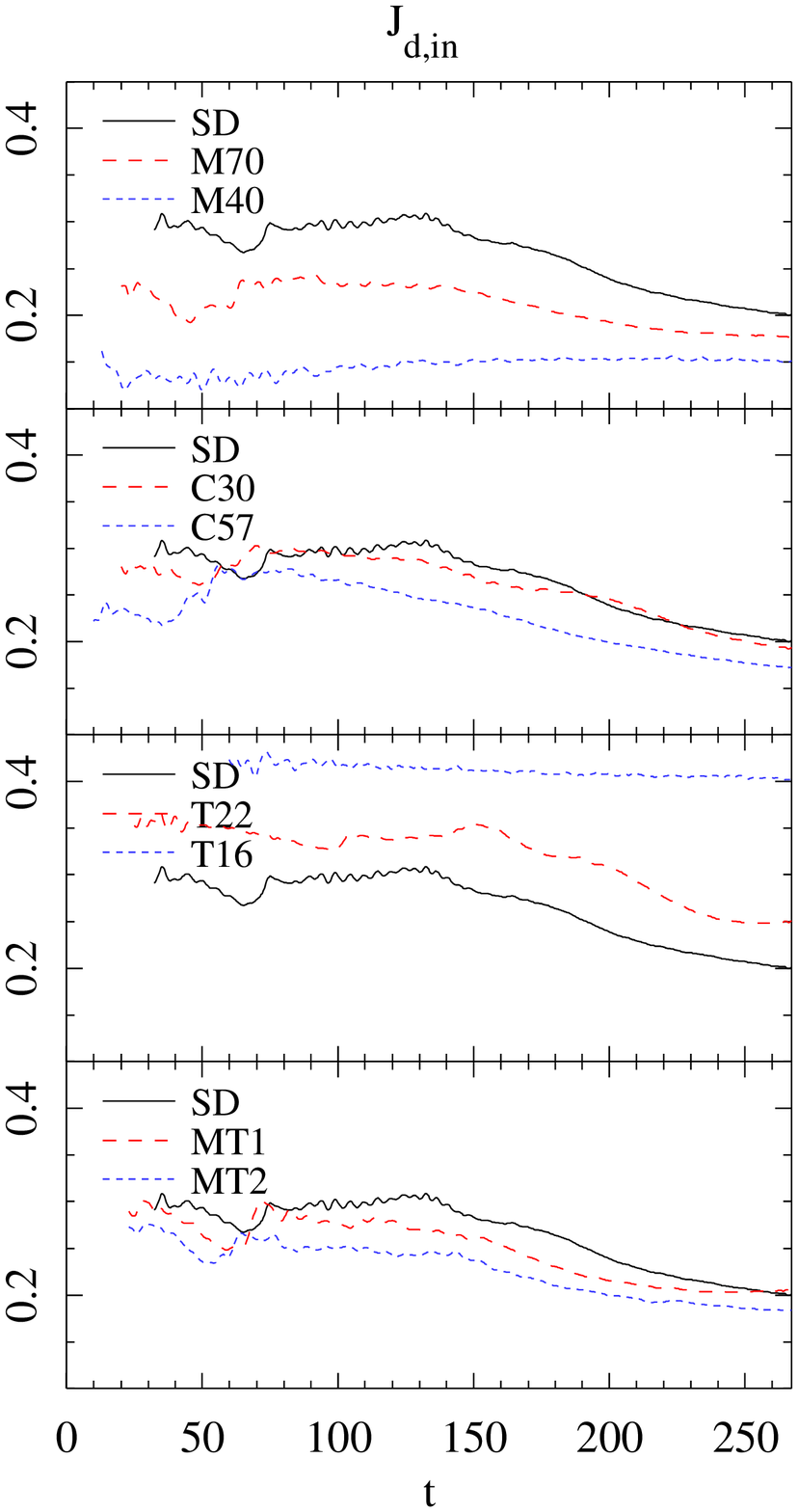}%
   \hfill\includegraphics[width=0.3\linewidth,bb=115 41 463 716]{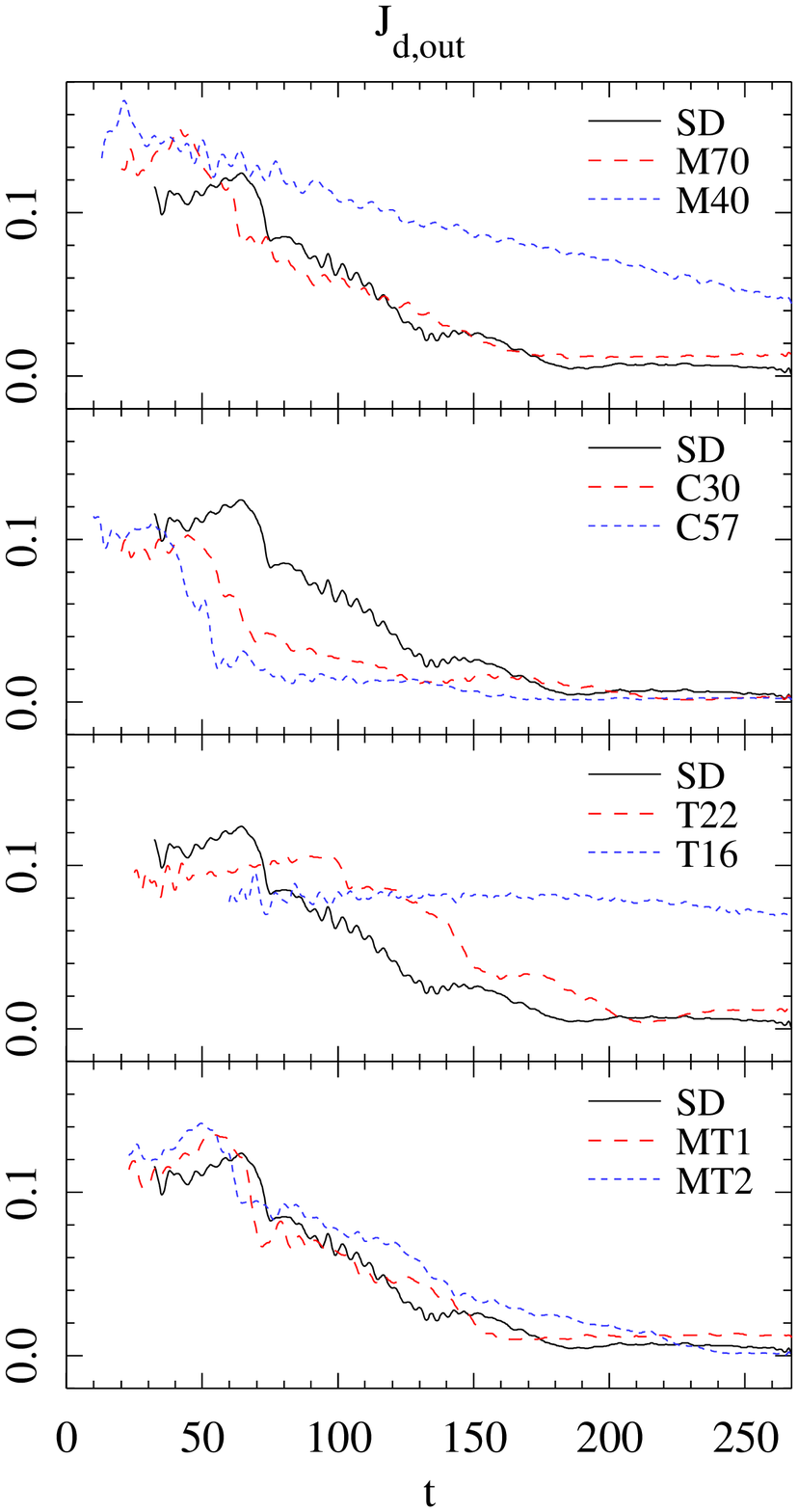}
   \caption{Evolution of the angular momentum, \Jd{},
        in the disk enclosed by the cylindrical volumes: $R\leq R_{\rm d}$
        (left), by $R\leq R_{\rm cr}$ (center), \Rcr$\leq R \leq 
        R_{\rm d}$ (right). The vertical extent of integration is taken
        as $z=\pm 0.3$. Note that the vertical scales are different 
        in each column.}
   \label{Jzd}
\end{figure*}

\subsection{Evolution of the angular momentum}

\begin{figure}
   \centering
    \includegraphics[width=1.0\linewidth,bb=52 300 370 675]{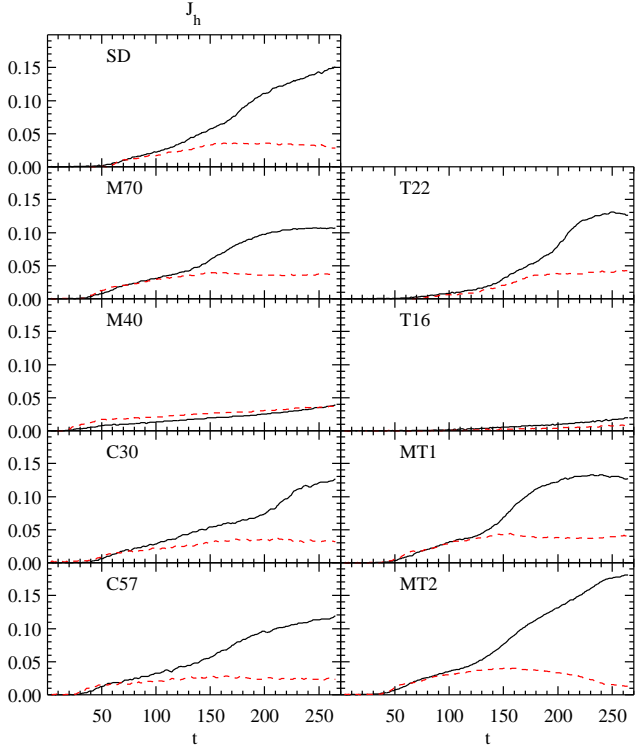}
   \caption{Evolution of the $z$ components of the halo angular momenta, \Jhin{}  
        (solid, black) and \Jhout{} (dashed, green), enclosed by the cylindrical
          volumes $R\leq $ \Rcr{} and $R\geq $ \Rcr{}, and $z=\pm 2$.
}
   \label{Jzh}
\end{figure}

The backbone of the bar is composed from material moving along eccentric orbits 
of relatively low angular momentum, compared to the circular orbits of the
same energy. Moreover growing bars consume more circular orbits, i.e.,
those with larger $J/E$ ratio, redistributing
their angular momentum to the outer disk and halo.
The capacity of different regions of the 
disk and halo to emit and absorb angular momentum is thus of a prime 
importance to the bar evolution, and largely determine its ability to grow 
by trapping additional disk orbits. Rather than estimate the effect of the
resonances on this process (Athanassoula 2002; Martinez-Valpuesta et al. 2006),
we focus on the total $J$ transfer within the disk and to the halo, i.e.,
accounting for the resonant and non-resonant $J$ redistribution.
In Figure~\ref{Jzd} we plot the evolution of the $z$ component 
of the total angular momentum of the disk, \Jd{}, as well as in the
cylindrical volumes centered on the symmetry axis of the disk
and separated by \Rcr, i.e., $R \leq $\Rcr{} (\Jdin{}) and 
\Rcr{} $\leq R$ (\Jdout{}), and similarly for the halo (Fig.~\ref{Jzh}).
Since the total angular momentum in each model is conserved, the evolution 
of \Jh{} in the halo is an inverted mirror image of that in the disk.

The initial angular momentum, \Jd{($t=0$)}, of the disk varies from model 
to model due to the differences in the halo mass distribution. It changes 
very little before the vertical buckling in the bar, while \Jdin{} and 
\Jdout{} anticorrelate. This means that the halo gains a negligible amount 
of $J$, which flows nearly exclusively to the outer disk during this 
phase (see also Fig.~\ref{Jzh}). After the buckling, \Jd{} and \Jdout{} 
drop monotonically till the end of the run. This is not true about 
\Jdin{} which stays approximately constant 
until the time when \Rcr{} leaves the disk (Fig.~\ref{Jzd}; see also 
Martinez-Valpuesta et al. 2006). How this is possible?

While it is not surprising that the loss of $J$ within the corotation
leads to an outward motion of \Rcr, the apparent `conspiracy' in
\Jdin{}$\sim const.$ over such an extended time period is somewhat
puzzling. It hints at some regulation mechanism which
fine tunes the loss of angular momentum to the outer disk and to the
halo by gravitational torques with its influx across the corotation.
We attempt to answer this question, using obvious simplifications:
the explanation of this phenomenon lies in the intricacies of \Jdin{} 
balance and the resulting behavior of \ps{}. The angular momentum
within \Rcr{} is
\begin{equation}
\label{eq:Jd}
J_{\rm d,in} = 2\pi \int_0^{R_{\rm cr}(t)} dR R^2 \Sigma(R,t) \bar{v}_{\rm t} ,
\end{equation}
where $\bar{v}_{\rm t}\equiv \alpha v_{\rm c}$ is the average tangential
velocity of particles inside \Rcr{}, which also defines the coefficient $\alpha$.
We require \Jdin{}$\sim const.$ between the times of buckling\footnote{In 
fact, we can impose this condition already from $t=0$, as variation of 
\Jdin{} is relatively small during the buckling (e.g., Fig.~\ref{Jzh})}
and when \Rcr$/R_{\rm d}\sim 1$, and take the time derivative of 
\Jdin{} using the Leibniz formula,
\begin{eqnarray}
\label{eq:derivJd}
\frac{1}{2\pi}\frac{d J_{\rm d,in}}{dt} = 
     \int_0^{R_{\rm cr}} R^2
     \frac{d}{dt}[\Sigma(R,t) \bar{v}_{\rm t}] dR + 
     \nonumber\\
     + v_{\rm c} R_{\rm cr}^2 \Sigma(R_{\rm cr},t)
     \frac{dR_{\rm cr}}{dt} = 0,
\end{eqnarray}
where $\Sigma$ is the disk surface density which can be obtained from 
eq.~\ref{eq:rho_disk}
by integrating over $z$. We assume $v_{\rm c}\sim const.$ This is justified
because contribution of $J$ within the velocity turnover radius is
small and at larger radii, $v_{\rm c}$ is dominated by the isothermal halo
and is nearly independent of radius (Fig.~\ref{vcir3}). 
The first term in eq.~\ref{eq:derivJd} describes the emission of $J$ by the 
{\it inner} disk, 
within an instantaneous \Rcr, due to the torque $T$ imposed by the DM and the outer
disk. For brevity, we assume that the stellar bar extends to \Rcr{} and this
torque acts upon it. In other words, all the material within \Rcr{} resides in the
bar. The second term in eq.~\ref{eq:derivJd} represents the
influx of \Jdin{} due to the advance of \Rcr{} and the resulting addition of a new 
mass within this radius on nearly circular orbits. Next, we relate the
motion of \Rcr{} to the slowdown of the bar, $dR_{\rm cr}/dt\equiv
\dot R_{\rm cr} = - R_{\rm cr}{\dot \Omega}_{\rm b}/\Omega_{\rm b}$.
The torque on the bar
gives the rate of change of its angular momentum (and, therefore, change in the
angular momentum of the inner disk):
\begin{equation}
\label{eq:torque}
T = \frac{d}{dt} (I_{\rm b}\Omega_{\rm b} + J_{\rm circ}) ,
\end{equation}
where $I_{\rm b}$ is the moment of inertia of the bar and $J_{\rm circ}$ is the
angular momentum of the internal circulation within the bar. (The second term does
not appear in eq.~10 of Athanassoula (2003).) 
Hence, $T=\dot \Omega_{\rm b} I_{\rm b} + 
\Omega_{\rm b} \dot I_{\rm b} + \dot J_{\rm circ}$. 
Assuming that the dominant loss of the angular momentum by the bar is due 
to the slowdown of its tumbling 
$\dot \Omega_{\rm b}$ (not a trivial assumption and definitely not a general one!),
we can re-write eq.~\ref{eq:derivJd}, replacing its first term by the action of $T$, 
as
\begin{equation}
\label{eq:torque2}
\dot \Omega_{\rm b} I_{\rm b} - v_{\rm c} R_{\rm cr}^3 \Sigma_{\rm cr}
     \frac{\dot\Omega_{\rm b}}{\Omega_{\rm b}}\approx \dot\Omega_{\rm b}(I_{\rm b} -
       R_{\rm cr}^2 M_{\rm b}) = 0 ,
\end{equation}
where we have taken $M_{\rm b}\sim R_{\rm cr}^2\Sigma_{\rm cr}$. If the
second term in eq.~\ref{eq:torque2} loosely represents the moment of inertia
of the bar, the net change in \Jdin{} is indeed negligible, explaining its
near constancy in Fig.~\ref{Jzd} (mid column) up to the time when \Rcr{} leaves the 
disk. In summary, we show that the constancy of the angular momentum within \Rcr{} 
can be
indeed explained if a number of straightforward assumptions is made to allow for an analytical
estimates. These assumptions are as following: (1) the bar extends to near \Rcr{} 
during this time and the disk mass within this radius lies mainly in the bar; (2) the 
rate of change of the angular momentum in the bar due to the change in its 
moment of inertia and internal circulation is smaller than that resulting from the 
bar slowdown; (3) the circular velocity in the disk around the \Rcr{} region and 
beyond is independent of radius; and (4) the average tangential velocity of the material in the bar is a fixed fraction of the disk circular velocity.
While this behavior shows up in all our models, as long as \Rcr{} lies 
within the disk, a more general set of mass distributions should be tested 
before making a final conclusion.

\begin{figure}[ht!!!!!!!!!!!]
   \centering
\includegraphics[width=0.65\linewidth]{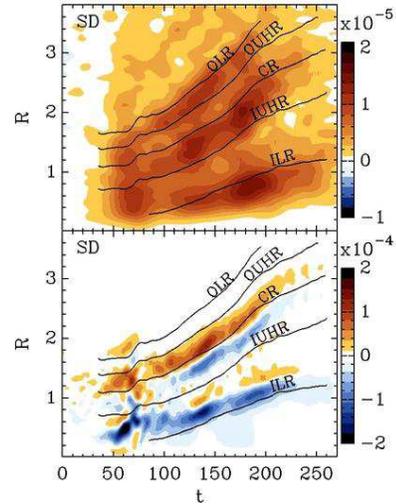}
   \caption{The rate of angular momentum flow \dJ{} as a function of radius 
        and time for the SD model (see also the text). 
	The top row corresponds to the halo and the bottom row to the disk.
        Red/blue colors represent the absorption/emission of $J$ using a
        linear scale in color. The main
        resonances are indicated by the solid lines: the outer/inner 
        Lindblad resonances (OLR/ILR), the outer/inner ultraharmonic 
        resonances (OUHR/IUHR) and the corotation (CR). The
        cylindrical shells have $\Delta R = 0.1$ and $z=\pm 2$. Time
        smearing is $\Delta t = 3$.}
   \label{Jz_diag_SD}
\end{figure}

While the halo absorbs all $J$ emitted by the disk, it is instructive
to look into which parts of the halo are especially active in this 
process. Figure~\ref{Jzh} displays the evolution of the absorbed
\Jhin{} and \Jhout. We observe that the inner halo becomes more
receptive to the process with time, except in M40 and T16. This
is clearly related to the outward motion of \Rcr{}. We note also,
that the rate of $J$ absorption by the inner halo increases
substantially after \Rcr/$R_{\rm d}\sim 1$, and the halo serves
as the only sink of the angular momentum. Asymptotically, one
can divide the models in two classes: when the rate of $J$ absorption
by the inner halo goes to zero or even becomes negative (e.g., M70, T22,
MT1), or when it declines only slightly (e.g., SD, M40, C30). 
 
\begin{figure}[!ht]
   \centering
    \includegraphics[width=\linewidth]{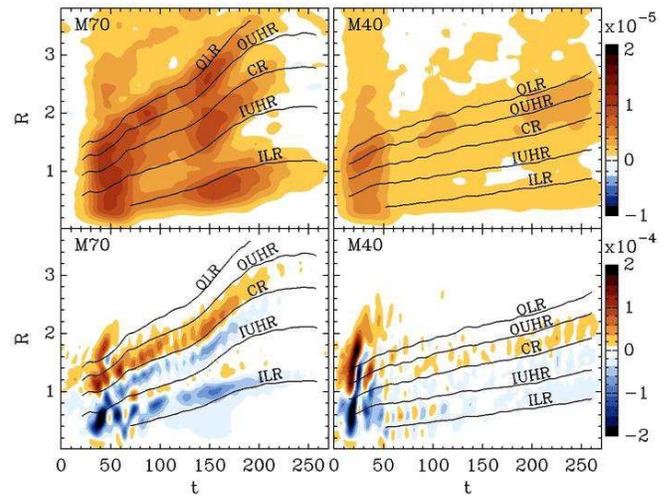}
   \caption{\dJ{} as a function of radius and time for models M70 (left 
        column) and M40 (right column).
	The top row corresponds to the halo and the bottom row to the disk.
        Color code and lines as in Fig.~\ref{Jz_diag_SD}.}
   \label{Jz_diag_M}
\end{figure}

\begin{figure}[!ht]
   \centering
    \includegraphics[width=\linewidth]{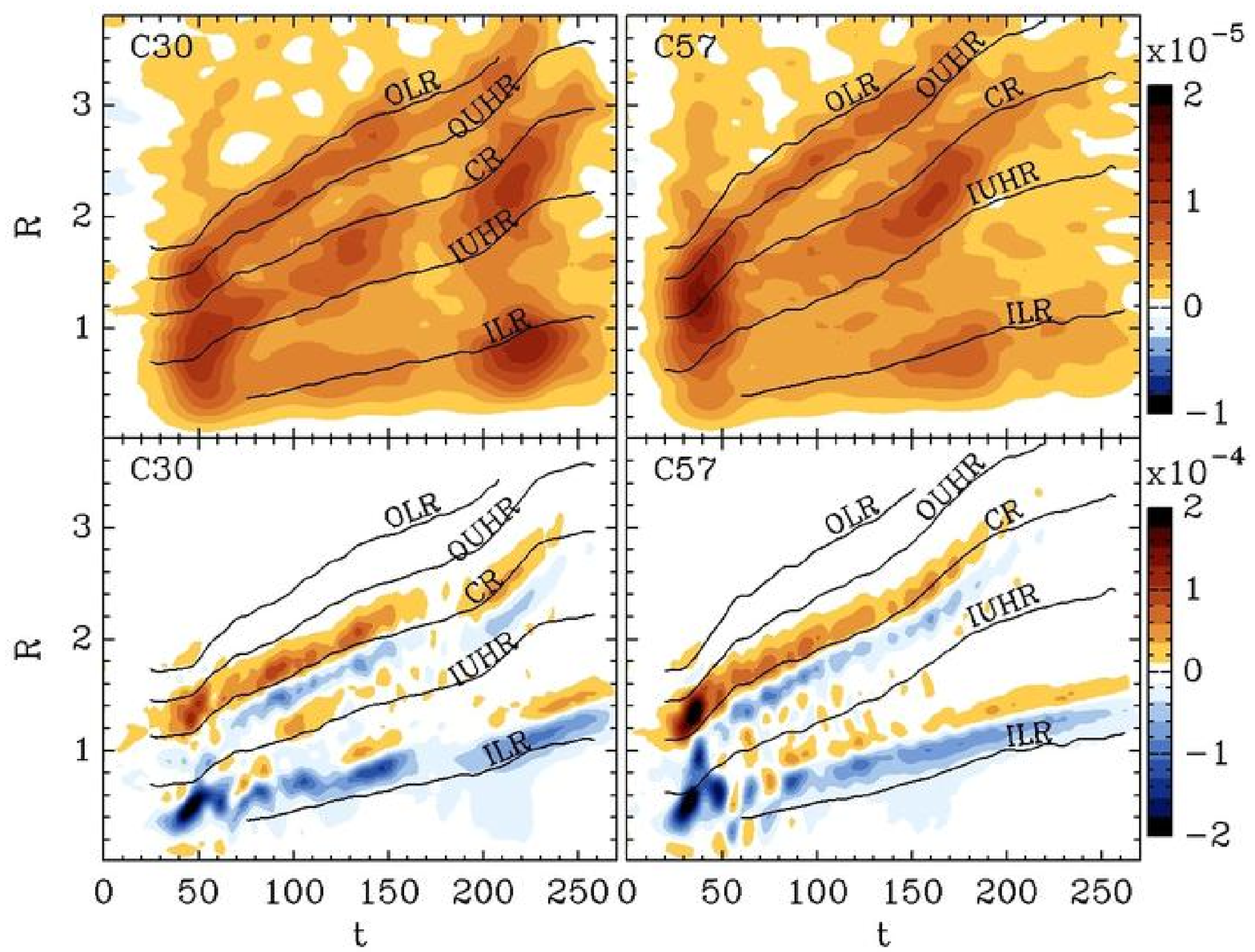}
   \caption{\dJ{} as a function of radius and time for models 
        C30 (left column) and C57 (right column). 
	The top row corresponds to the halo and the bottom row to the disk.
        Color code and lines as in Fig.~\ref{Jz_diag_SD}.}
   \label{Jz_diag_C}
\end{figure}

\begin{figure}[ht!!!!!!]
   \centering
   \includegraphics[width=\linewidth]{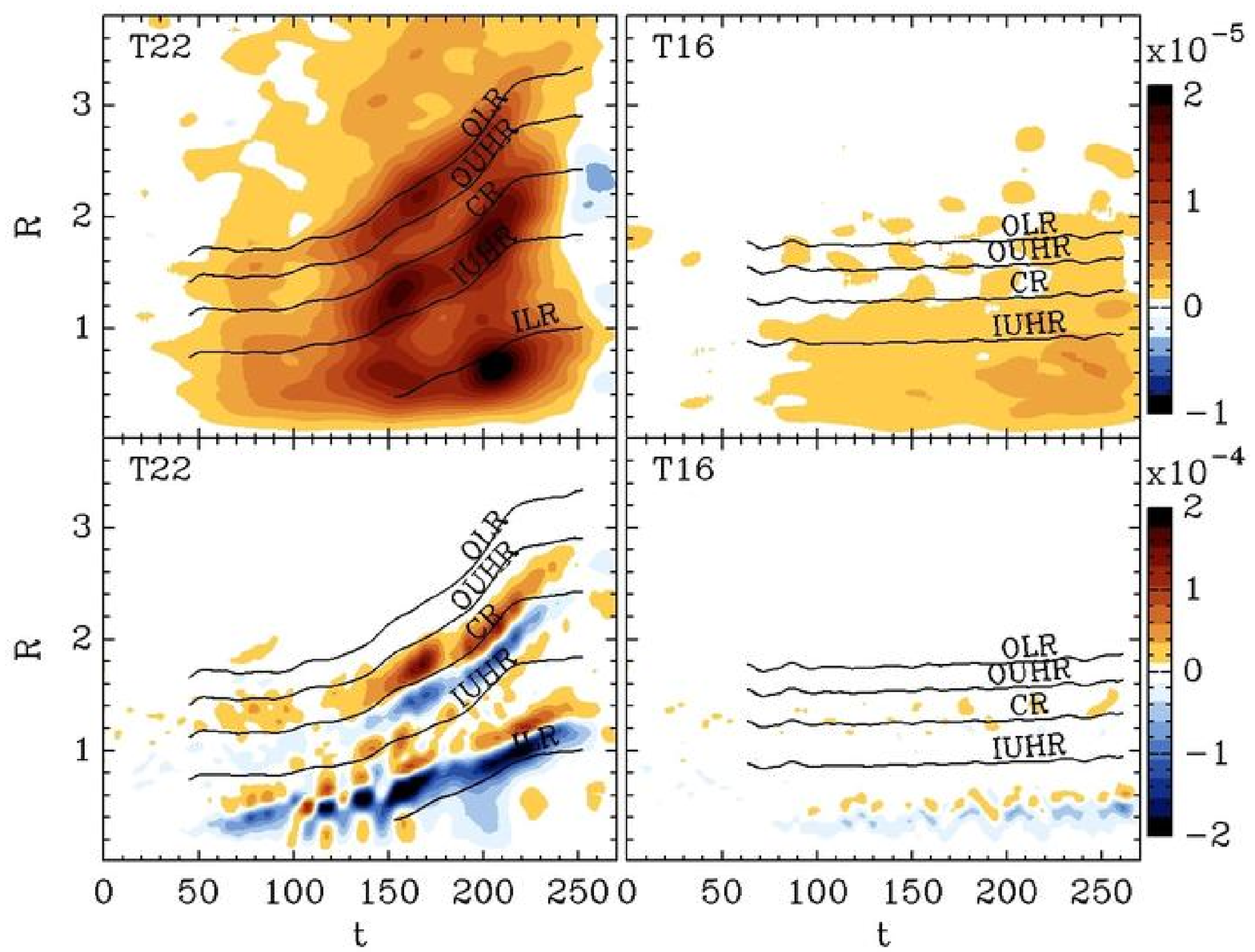}
   \caption{\dJ{} as a function of radius and time for models 
        T22 (left column) and T16 (rigt column). 
	The top row corresponds to the halo and the bottom row to the disk.
        Color code and lines as in Fig.~\ref{Jz_diag_SD}.}
   \label{Jz_diag_T}
\end{figure}

\begin{figure}[!ht]
   \centering
    \includegraphics[width=\linewidth]{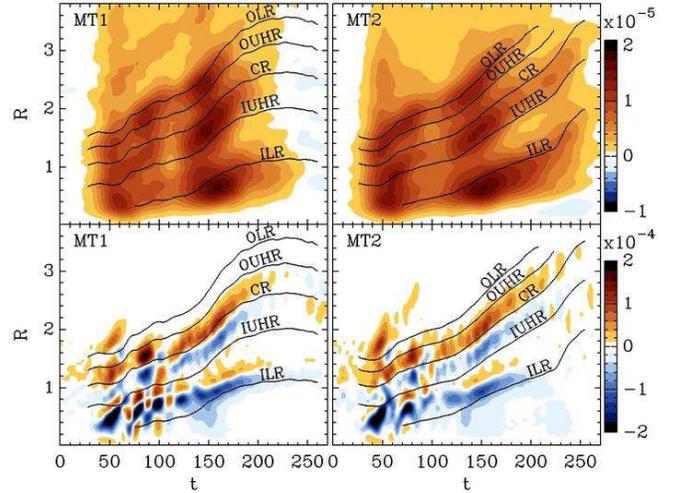}
   \caption{\dJ{} as a function of radius and time for models
        MT1 (left column) and MT2 (right column). 
	The top row corresponds to the halo and the bottom row to the disk.
        Color code and lines as in Fig.~\ref{Jz_diag_SD}.}
   \label{Jz_diag_MT}
\end{figure}

An alternative way to detail the exchange of the angular momentum 
in the disk-halo system is to divide the disk and halo into a number
of concentric cylindrical shells. We have constructed a 2-D map of 
the angular momentum in each shell as a function of radius and time. 
It is more revealing, however, to plot the time derivative of \Jd{}
and \Jh{} at each radius. This analysis was performed separately for 
the halo and the disk. In Figures~\ref{Jz_diag_SD} to \ref{Jz_diag_MT}, 
we display color-coded diagrams of the rate of change of angular momentum 
$\dot{J}_{d,R}=(\partial J_{\rm d}/\partial t)_{\rm R}$ with radius and 
time for each run, and the same for the halo. Red color 
corresponds to emission and blue to absorption of the angular momentum.
All the diagrams corresponding to disks have been calibrated identically. 
DM halos have also been calibrated uniformly among themselves. 

The most striking features in Figures~\ref{Jz_diag_SD} to \ref{Jz_diag_MT}
is the continuity of color, i.e., emission and absorption bands, both 
in the disk 
and the halo, and their outward shift. The bands appear nonuniform in time 
and space. The halos are almost exclusively absorbers of the angular momentum, 
while the disks show regions of absorption and emission. 
From the halo evolution in the SD model (Fig.~\ref{Jz_diag_M}), we note that 
\Jh{} is absorbed preferentially at three distinct bands which shift outwards 
with time. Other and weaker absorbing regions are observed between them and
at larger radii. The absorption/emission bands in the disk are alternating,
unlike in the halo.
 
To relate the emission/absorption bands to dynamical characteristics of the
disk-halo system, we specify the positions of
linear resonances on the color map. These are used for guidance purpose
only, as the positions of linear resonances maybe misleading, especially
for the strong bars, and especially for the inner Lindblad resonance(s). 
 
We find a close correspondence between the behavior of \dJd{(R)} 
bands and the location of the linear resonances (superposed solid lines), 
namely,
the corotation, inner/outer Lindblad resonances (ILR/OLR), and the 
inner/outer ultra-harmonic resonances (UHR). While it is clear that the
redistribution of $J$ in the disk and the halo is a strongly nonlinear
process, some general conclusions can be made. 

First, as expected, the emission and absorption between the inner disk 
and the inner halo components strongly correlate, with a certain interference 
of the outer disk (Figs.~\ref{Jz_diag_M}--\ref{Jz_diag_MT}). Here we 
refer to inner/outer with respect to the
\Rcr. The peak emission of $J$ by the disk happens at the band associated
with the ILR. A weaker emission band lies at the inner UHR (see also
Fig.~10 of Martinez-Valpuesta et al. 2006). The strongest
absorption in the disk lies at the corotation and extends to the outer
UHR. Weaker absorption is associated with the OLR in the disk. The halo
absorption bands are centered on the ILR, inner UHR/corotation and
the outer UHR/OLR. However, it is nonnegligible in other regions as well.
We note that producing horizontal slices in the halo reveals that
the lower halos, i.e., $\Delta z=\pm 0.3$, absorb mostly at the ILR while
the upper halo absorb mostly at \Rcr. This is a refinement to previous
results which shown that the main absorption is at the CR resonance 
(Athanassoula 2002; Martinez-Valpuesta et al. 2006).

Second, we observe temporal correlations between the emission and absorption
peaks in the disk and the halo. Over the length of the run, both the
disk and the halo experience from two to three `times of activity,'
except T160. The first such activity time is clearly associated with the 
first peak in \Atwo{} and the subsequent buckling. The other activity time
lies toward the saturation of \Atwo{}, late in the evolution of the bar. 
Some models, like SD, show an additional activity in between. This seems
to be associated with \Rcr{} crossing $R_{\rm d}$ --- the halo absorption
peaks at the corotation during this time. The resonances move 
outwards faster during intense $J$ absorption. The outward migration 
of the resonances is of course related to the bar slowdown, resulting in 
the resonances sweeping across the disk and halo material.  

Third, \dJd(R) and \dJh(R) in Figs.~\ref{Jz_diag_M}--\ref{Jz_diag_MT} 
appear to correlate with \Jdin{}, \Jdout{}, \Jhin{} and \Jhout{}
in Figs.~\ref{Jzd} and \ref{Jzh}. It also correlates with
\Atwo{} and especially with \Atworb{}. Lastly, the maxima of disk
emission correspond to the maxima in the halo absorption.
 
Next, we return to the atypical behavior of some stellar bars in their
secular phase, which exhibit \ps$\approx const.$, discussed in \S3.1.
In all the models where we observe this phenomenon, the ratio 
\Rcr/$R_{\rm d} > 1$ at the time of detection. This behavior, as expected,
is associated with the rate of $J$ transfer from the disk to the halo:
in all these models, $\dot J\rightarrow 0$, both in the halo absorption
or disk emission during $\Omega_{\rm b}\approx const.$ 
(Figs.~\ref{Jz_diag_M} to \ref{Jz_diag_MT}). The outward motion of the
disk resonances stalls as well (Figs.~\ref{Rcd} and \ref{Jz_diag_M} --- 
\ref{Jz_diag_MT}). Moreover, there is an
indication that this is concurrent with the reversal of $J$ flow:
while the disk still shows a weak emission at its ILR, the halo
shows a weak {\it emission} at \Rcr{} and a strong absorption at the ILR. 
The latter remains the most active resonance in the halo. The emission
of $J$ by the halo at \Rcr{} is rather unusual and has not been observed 
before in any models in the literature, to the best of our knowledge.

Lastly, the disk inspection reveals the presence of two trailing spiral 
arms extending to \Rcr{}, and when \Rcr/$R_{\rm d} > 1$, connecting the
outer edge with \Rcr{} (see \S3.3). Some stellar particles appear to be
trapped at \Rcr{} in its outward motion up to $t\sim 250$, depending
on the model. These particles appear to concentrate at specific azimuthal
locations of the CR ``circle'' with respect to the bar (in the bar frame 
of reference). 

The particles in the spirals are not trapped.  These spirals are prominent 
in Figs.~\ref{Jz_diag_M}--\ref{Jz_diag_MT}, and are responsible for
emission at \Rcr{} and absorption of $J$ in the IUHR-CR band, as \Rcr{}
moves outside the disk. While the number of these particles is not large,
their specific angular momenta is the largest in the disk.

\section{Discussion: Bar-Disk-Halo Correlations}
\label{s_disc}

We have constructed and evolved three one-parameter and one
two-parameter sequences of DM halos hosting a standard asymmetric stellar 
disk. All models developed bars and we have followed their evolution
over a Hubble time. The properties of the bars, such as their strength,
pattern speed, size, corotation radius, ratios of corotation to bar lengths,
and more, have been related to those of the disks and host halos. 
Based on our results, we now test a number of correlations found by
Athanassoula (2003). We also obtain additional ones --- all of these will
be used in order to compare with the gas models (Paper~II). 

For this purpose, we deal with three entities: the inner, bar-forming disk 
(defined within \Rcr), the disk as a whole, and the DM halo. We 
adopt the halo-to-disk mass ratio, $M_{\rm h}/M_{\rm d}$ within the disk 
velocity turnover radius, $R_{\rm turn} = 2.2h$ (e.g., Sackett 1997) at 
$t=0$, as an independent variable, where $h$ is the disk radial scalelength 
defined in section~\ref{s_inicon}. In the following, we also use the
radial and vertical dispersion velocities in the disk and radial 
dispersions in the halo, and test how the efficiency of $J$ 
redistribution within the system depends on these parameters.  

\begin{figure}[!ht]
   \centering
    \includegraphics[width=0.9615\linewidth]{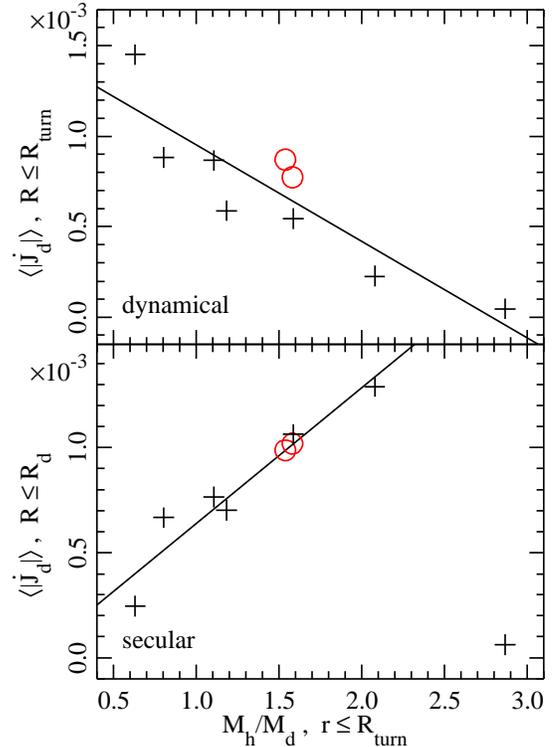}
   \caption{Average rate of angular momentum loss, $\langle|\dot J|\rangle$, 
       by the 
       inner stellar disk during its dynamical (upper) and secular (lower)
       phases of the bar evolution as a 
       function of the $M_{\rm h}/M_{\rm d}$ ratio. The rate $\langle\dot J\rangle$
       has been calculated within $R_{\rm turn}$ (upper) and
       within $R_{\rm d}$ (lower). The 
       mass ratio has been calculated within $R_{\rm turn}$. The
       averaging was performed over the time period from $t=0$ to the
       end of an exponential growth of the stellar bars (upper) and from
       the time of the first minimum of \Atwo{}, i.e., the of the buckling, 
       to its saturation, $t=213$
       (about 10~Gyr) (lower). The green circles represent
       the two hybrid models. The attempted linear fit in the lower frame does
       not include the lower right corner point (T16).
}
   \label{dJMhMd_in}
\end{figure}

To quantify the role of the halo in the bar instability, we calculated the
exponential timescale of \Atwo{} growth, $t_{\rm bar}$, against
$M_{\rm h}/M_{\rm d}$ within $R_{\rm turn}$, which represents the halo 
concentration, normalized by the disk mass within the same radius. A linear 
correlation was obtained, namely, that an increase in the DM mass concentration
leads to a slower bar growth before the first buckling. However, the
bar evolution in the secular phase cannot be described by an exponential
growth. Looking for a more
universal measure of the bar growth at all times, we decided in favor of
angular momentum change in the inner and full disks. Figure~\ref{dJMhMd_in} 
displays the average rate of angular
momentum change, $\langle |\dot J|\rangle $, in the dynamical and secular 
phases of the bar evolution, within 
$R_{\rm turn}$ and $R_{\rm d}$ respectively. The former choice of a fixed 
radius ($R_{\rm turn}$) results from
total \Jd{} being nearly unchanged before the buckling --- $J$ is exchanged
predominantly between the inner and outer disks across \Rcr{} at this stage. 
We have refrained from using \Rcr{} as an inner/outer disk
separator here because its outward motion brings in fresh, high $J/E$
material which `contaminates' the $J$ transfer due to the resonant and
non-resonant interactions between the inner and outer disks.
The latter choice of $R_{\rm d}$ in the secular phase results from \Jdin{} 
being nearly constant --- $J$ is mostly exchanged between the disk and the 
halo. 
 
\begin{figure}[!!!!!!!!!!!ht]
   \centering
 \includegraphics[width=0.9615\linewidth]{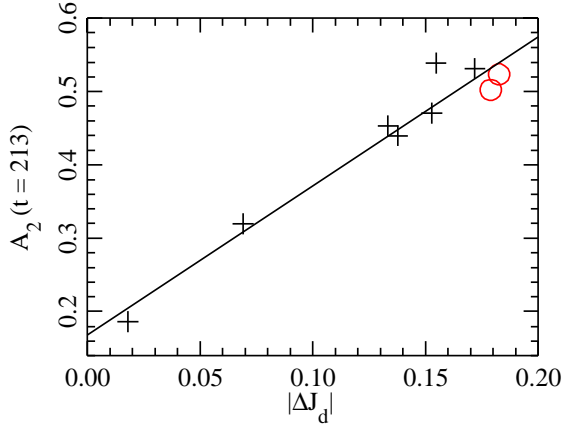}
   \caption{\Atwo{}\ at $t\sim 213$, corresponding to 10~Gyr,  as a function 
    of the angular momentum change in the whole disk 
    over $\Delta t = 213$. The green circles represent the two hybrid models.}
   \label{A2vsdJ}
\end{figure}

Figure~\ref{dJMhMd_in} shows that clear correlations of  
$\langle |\dot J|\rangle $ with 
$M_{\rm h}/M_{\rm d}$ persist over the Hubble time --- more concentrated 
halos provide an increasingly efficient damping of the bar instability 
by reducing the rate of $J$ transfer to the outer disk (upper frame). They 
do facilitate such a transfer
in the secular phase (lower frame). Significantly, these correlations extend 
down at least 
to $M_{\rm h}/M_{\rm d}\sim 0.5$ --- the disk-dominated models. We note
that T16 exhibits a qualitatively and quantitatively different behavior
and was not included in the attempted linear fit to the secular evolution
(lower frame). We have discussed this model in \S3.4.

The explanation for such a dual role of a DM halo in the bar evolution lies 
in determining what serves as a sink of \Jdin{} from the inner disk. As shown 
in \S3.6, 
the outer disk beyond \Rcr{} absorbs nearly all of $J$ during the bar 
instability. A more concentrated halo during this phase will dilute the
disk gravity and, therefore, will act against the resonant orbit coupling 
(i.e., resonant torques) between
the inner and outer disks. On the other hand, during the secular phase
of the bar instability, after the first buckling, the halo serves as
the sink of $J$ from the disk. 
In this case, if increase of the DM mass density leads to a concurrent 
increase in the DM phase space density near the resonances, it facilitates 
the bar-halo resonance coupling, unless counterbalanced 
by a `hotter' halo. The $\langle\dot J\rangle$ rates appear similar
during the dynamical and secular phases. But the duration of each 
phase can differ considerably. Therefore, one expects that the amount of $J$ 
acquired by the halo will be much more substantial during the secular phase
compared to the angular momentum lost by the inner disk (within a fixed
radius!) during the dynamical phase.

It is known that the DM halo affects the bar evolution. In particular, it
was shown that the DM halo concentration anti-correlates with the bar
growth time. Athanassoula (2003) has also demonstrated the 
(anti)-correlation between $\Omega_\mathrm{b}$ and \Atwo{}. Our
Figure~\ref{dJMhMd_in} quantifies the dual role played by the DM halo during
the bar evolution and ties this explicitly to the rate of the angular momentum
transfer within the disk--halo dynamical system.

\begin{figure}[!ht]
   \centering
 \includegraphics[width=0.93\linewidth]{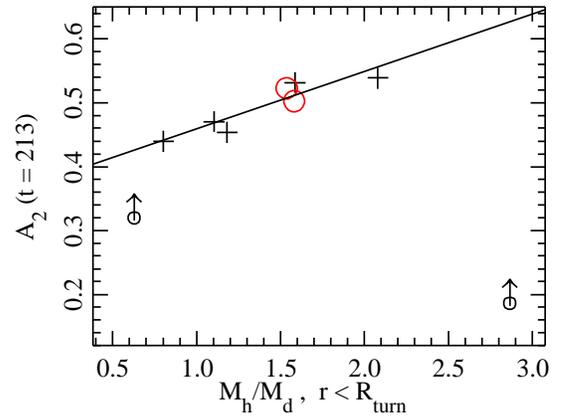}
   \caption{Saturated \Atwo{}\ at $t=213$ ($\sim 10$~Gyr), as a function of 
         the halo-to-disk mass ratio within $R_{\rm turn}$ at $t=0$. Two models
         do not reach the saturation and are shown as lower limits: M40
          (left) and T16 (right). The green circles represent the two hybrid   
          models.}
   \label{A2_hmassin}
\end{figure}
\begin{figure}[!ht]
   \centering
   \includegraphics[width=0.85\linewidth]{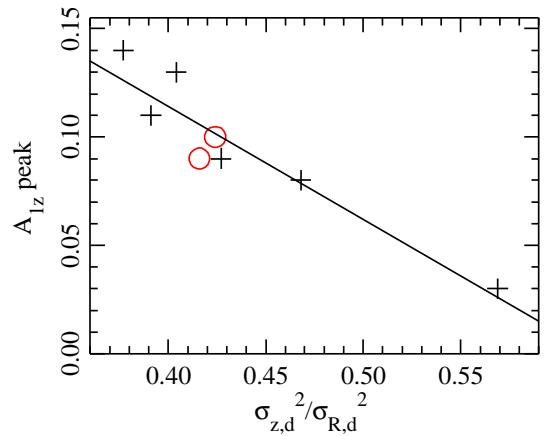}
   \caption{Dependence of \Aone{} amplitude on the ratio of 
           vertical-to-radial velocity dispersions, 
           $(\sigma_{\rm z,d}/\sigma_{\rm r,d})^2$. T16 is excluded as it 
           does not buckle. The green circles represent the 
           two hybrid models. These velocity dispersions have been measured 
           in a ring defined by $0.1 \leq R \leq 0.2$ and $|z| \leq 0.1$.}
   \label{A1_sigzsigr}
\end{figure}
\begin{figure}[!ht]
   \centering
    \includegraphics[width=0.96\linewidth]{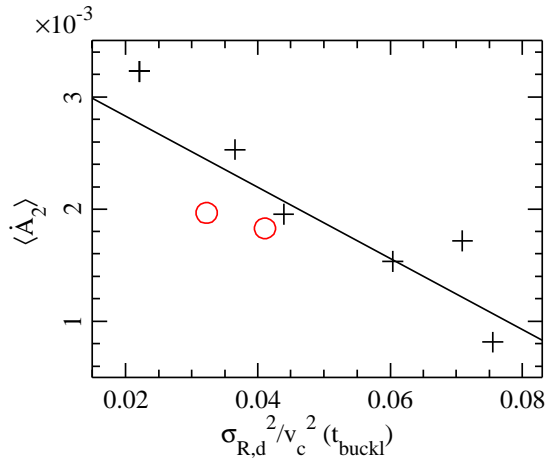}
   \caption{Secular evolution of the bars: dependence of 
        $\langle\dot A_2\rangle$, the average growth rate of the bar, between
        $t_{\rm buckl}$ and $t=213$ (10~Gyr), vs 
        $\sigma_{\rm r,d}/v_{\rm c}$, the ratio of the radial dispersion 
        in the disk to the circular velocity at $R_{\rm turn}$. The latter 
        ratio is calculated at $t_{\rm buckl}$. T16 is omitted as it does 
        not buckle. The green circles represent the two hybrid models.
}
\label{A2deriv_sig_vc}
\end{figure}
\begin{figure}[!ht]
   \centering
   \includegraphics[width=0.92\linewidth]{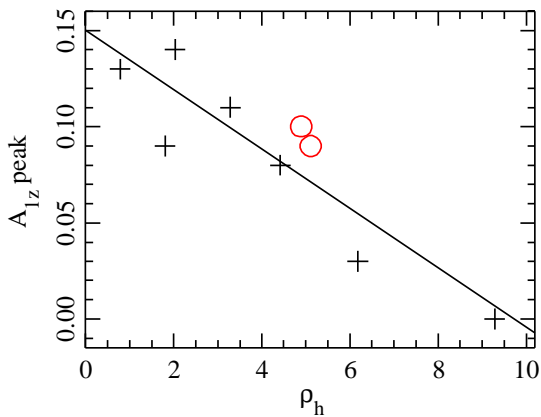}
   \caption{Dependence of  
        \Aone{} peak on the central halo density at $t=0$. The green 
        circles represent the two hybrid models.
        }
   \label{A1_hdens}
\end{figure}

Stronger bars are expected to be more efficient in 
redistribution of the angular momentum in the disk-halo systems. In our 
{\it collisionless} models, bars strengthen monotonically only prior to the
first peak and after the first buckling. Despite the overall non-monotonic
behavior of the bar, its final strength correlates with the amount of
$\Delta$\Jd{} lost by the disk and acquired by the halo (Fig.~\ref{A2vsdJ}),
in agreement with Athanassoula (2003).
 
For the same reason,
we expect and obtain correlation between \Atwo{} at $t=213$ ($\sim 10$~Gyr),
and $M_{\rm h}/M_{\rm d}$ ratio within $R_{\rm turn}$, i.e., the halo 
mass concentration (Fig.~\ref{A2_hmassin}), thus confirming Athanassoula 
(2003) claim. By this time, nearly all modeled bar strengths reach saturation
(e.g., Fig.~\ref{Rb_A2}), except in two models, M40 and T16, shown as lower 
limits. As shown in \S3.3, \Atwo{} and \Atworb{} saturate shortly after
\Rcr/$R_{\rm d}\sim 1$. These
models represent two extreme trends in the halo mass concentration. T16
has the most concentrated halo, does not buckle and its
secular evolution differs from other bars. M40 represents
the least concentrated halo with a fractionally most massive disk 
(Figs.~\ref{rhomods} and \ref{vcir3}). It has the shortest
exponential growth for the bar, $t_{\rm bar}\sim 10$. Therefore, it is 
puzzling why its bar strength does not saturate. While M40 
exhibits buckling and subsequent drop
in \Atwo{} and \Atworb{}, its secular growth is the slowest of all models
and comparable to that of T16. Its bar is the longest of all 
models as the ansae do not disappear.
Furthermore, its \Rcr{} lies always within the disk, while \Rcr/\Rbar
$\sim 1.2$ and remains flat for a Hubble time. All these properties 
appear unique in our models.
 
While the behavior of T16 is partially understandable, that of M40 
requires more explanation. It is difficult to explain within the context 
of the halo mass concentration only, and seems
to require additional parameter. We invoke the disk radial 
dispersion velocity, $\sigma_{\rm R,d}$, which in a way also governs
the bar vertical buckling instability through its ratio to the vertical
velocity dispersion, $\sigma_{\rm z,d}$. We test this idea by first 
plotting the amplitude of buckling, \Aone, vs this ratio calculated just 
before the first buckling (Fig.~\ref{A1_sigzsigr}). Indeed, M40 has the 
smallest 
$(\sigma_{\rm z,d}/\sigma_{\rm R,d})^2$ among our models, and, therefore, 
is expected and, in fact, displays the largest buckling amplitude 
\Aone{} (Fig.~\ref{Rb}). 
A side effect of this buckling is the overall heating of the disk. 
Consequently, the disk will be less susceptible to the bar growth in the 
secular phase. 
Fig.~\ref{A2deriv_sig_vc} shows the growthrate of the bar strength
averaged over the secular phase, $\langle\dot A_2\rangle$, 
i.e., between the buckling time, $t_{\rm buckl}$, and the saturation 
time, $t\sim 213$ ($\sim 10$~Gyr), vs
$\sigma_{\rm R,d}/v_{\rm c}$ measured at $t_{\rm buckl}$ (i.e., at the 
minimum of \Atwo). A clear correlation
exists between these parameters --- `hotter' disk impairs the bar growth
in the secular phase of its evolution. Note that M40 has the hottest
disk in this stage, which would explain its anomalously slow bar growth.
The same correlation against $\sigma_{\rm R,d}$ 
looks weaker and shows much larger dispersion of individual models.

As mentioned before, in each separate sequence, the \Aone{} peak is 
stronger for models progressively less dominated by the halo.
It is interesting that an anticorrelation between the initial central 
density of the halo and the strength of the first \Aone{} peak holds for 
our models altogether, regardless of which halo parameter is 
varied (Figure \ref{A1_hdens}). 
Moreover, the anticorrelation exists
between the strength of the first \Aone{} peak and the ratio of the 
vertical-to-radial dispersion velocities, shown in Figure~\ref{A1_sigzsigr}.  
 
To summarize, we confirm the previous results and quantify for the first
time the dual role that the DM halos play 
in stellar bar evolution (Fig.~15): more centrally 
concentrated halos slowdown dynamical processes
in the disk, such as spontaneous bar instability and vertical buckling
instability, as well as angular momentum redistribution in the system. 
They reverse this trend and facilitate the angular momentum transfer during
the secular stage of bar evolution, following the buckling.  
We follow the angular transfer in the disk-halo systems by varying one basic
parameter at the time in order to identify the sites and times of intense
angular momentum flows. While we confirm the earlier works which have identified
the ILR and corotation resonances as being primarily responsible for $J$ 
emission by the disk and absorption by the halo, we also find few caveats.

First, the total angular momentum in the disk is largely conserved before
the first buckling, thus $J$ flows across the corotation to the outer disk,
although the amount is relatively low compared to the subsequent exchange.
Second, during the secular stage of the bar evolution, the angular momentum
within the corotation resonance is largely conserved: the loss of $J$
by this region due to the gravitational torques on the bar (resonant and
non-resonant) is compensated by the influx of new material, rich in $J$, 
across the corotation, due to its outward motion. We also elaborate under 
what conditions this `conspiracy' law operates. In view of this, the ratio
of \Rcr/$R_{\rm d}$ emerges as important dynamic discriminator between
various paths in barred disk evolution  --- its value can be determined 
if bar pattern speeds are known. Third, we find that in some
models the bar pattern speed stalls becoming nearly constant for prolonged
time periods. All stellar bars which show this behavior have their corotation
lying outside the disk. The disk-halo angular momentum exchange nearly vanishes
and the halos display a weak {\it emission} of $J$ at the corotation and
{\it absorption} at the ILR. While otherwise the bar pattern speeds are
generally strongly decaying over the secular timescale, the addition of the 
gas component can
modify this trend and soften the bar braking. We postpone our 
conclusions on this issue to Paper~II. We also find that \Rcr/\Rbar{}
ratio stays within 1---1.4 range, only occasionally spiking above it to
$\sim 1.8$. This behavior is typical after the first buckling and as long
as \Rcr/$R_{\rm d}\ltorder 1$. 

Lastly, we confirm some known correlations (Figs.~16 and 17) between the 
basic parameters of the
disk-halo system, e.g., between the final bar strength and $M_{\rm h}/M_{\rm d}$
ratio. Model M40 with the least concentrated halo does not follow this
trend and we test the possibility that an additional parameter plays the role
in this, i.e., the dispersion velocities ratio $\sigma_{\rm z,d}/\sigma_{\rm R,d}$.
Because M40 experiences the strongest buckling, the disk is heated up, which
impairs the subsequent bar growth and can explain the behavior of M40 in the
previous correlation. Figs.~18, 19 and 20 display new correlations between
various parameters. The bar average growthrate over its secular
evolution time decreases with increasing dispersion velocities in the disk,
$v_{\rm R,d}/v_{\rm c}$.
We also show that in a closely associated relation --- the angular 
momentum lost by the disk, $\Delta J_{\rm d}$, over the evolution correlates 
with the final bar strength.  Finally, the amplitude of the first buckling
depends on the central density in the DM halo.  These correlations will
be followed up when the gas component is present. We expect substantial
modifications in the relations between various basic parameters in the
disk-halo system.

\acknowledgments
We are grateful 
to Ingo Berentzen for comments on the manuscript, and to 
Ron Buta for answering our questions. We also thank Ingo Berentzen and Inma 
Martinez-Valpuesta for help with numerical issues.
This research has been partially supported by NASA/LTSA/ATP, NSF
and STScI grants to I.S. STScI is operated by the AURA, Inc., under NASA
contract NAS 5-26555. I.S. is grateful to the JILA Fellows for support.




\end{document}